\documentclass{aa}

\usepackage{graphicx}
\usepackage{txfonts}
\usepackage{courier}
\usepackage{color}
\usepackage{threeparttable}

\usepackage[compatibility=false]{caption}
\usepackage{subcaption}

\def\lognlogs{$\log(N)-\log(S)$}
\def\desqg{\deg^2}
\def\funits{\mathrm{erg}/\mathrm{cm}^2/\mathrm{s}}
\def\kev{\mathrm{keV}}

\def\atile{$68\arcmin\times 68\arcmin$}

\usepackage{natbib,twoopt}
\bibpunct{(}{)}{;}{a}{}{,}
\makeatletter
\newcommandtwoopt{\citeads}[3][][]{\href{http://adsabs.harvard.edu/abs/#3}%
{\def\hyper@linkstart##1##2{}
\let\hyper@linkend\@empty\citealp[#1][#2]{#3}}}
\newcommandtwoopt{\citepads}[3][][]{\href{http://adsabs.harvard.edu/abs/#3}%
{\def\hyper@linkstart##1##2{}
\let\hyper@linkend\@empty\citep[#1][#2]{#3}}}
\newcommandtwoopt{\citetads}[3][][]{\href{http://adsabs.harvard.edu/abs/#3}%
{\def\hyper@linkstart##1##2{}
\let\hyper@linkend\@empty\citet[#1][#2]{#3}}}
\newcommandtwoopt{\citeyearads}[3][][]
{\href{http://adsabs.harvard.edu/abs/#3}
{\def\hyper@linkstart##1##2{}
\let\hyper@linkend\@empty\citeyear[#1][#2]{#3}}}
\makeatother

\def\xamin{\texttt{XAmin}}
\def\xaminP{\texttt{XAminP06}}
\def\xaminF{\texttt{XAminF18}}
\def\sex{\texttt{SEXtractor}}

\begin{document}

\title{The XXL Survey XXIV. The final detection pipeline.}

\author{L. Faccioli     \inst{1} \and
        F. Pacaud       \inst{2} \and
        J.-L. Sauvageot \inst{1} \and
        M. Pierre       \inst{1} \and
        L. Chiappetti   \inst{3} \and
        N. Clerc        \inst{4} \and
        R. Gastaud      \inst{1} \and
        E. Koulouridis  \inst{1} \and
        A.M.C. Le Brun  \inst{1} \and
        A. Valotti      \inst{1}
}

\institute{
        AIM, CEA, CNRS, Universit\'{e} Paris-Saclay, Universit\'{e} Paris Diderot, Sorbonne Paris Cit\'{e}, \\
        F-91191 Gif-sur-Yvette, France \\
        \email{lorenzo.faccioli@cea.fr}
        \and
        Argelander Institut f\"{u}r Astronomie, Universit\"{a}t Bonn, 53121 Bonn, Germany
        \and
        INAF, IASF Milano, via Bassini 15, 20133 Milano, Italy
        \and
        IRAP, Universit\'{e} de Toulouse, CNRS, CNES, UPS, (Toulouse), France
}

\date{Received \today; accepted \today}

\abstract
{}
{A well characterised detection pipeline is an important ingredient for X-ray cluster surveys.}
{We present the final development of the XXL Survey pipeline.
The pipeline optimally uses X-ray information by combining many overlapping observations of
a source when possible, both for its detection and its characterisation.
It can robustly detect and characterise several types of X-ray
sources: AGNs (point-like), galaxy clusters (extended), galaxy clusters contaminated
by a central AGN, and pairs of AGNs close on the sky.
We perform a thorough suite of validation tests via realistic simulations of XMM-Newton images and
we introduce new selection criteria for various types of sources that will
be detected by the survey.}
{We find that the use of overlapping observations allows new clusters
to be securely identified that would be missed or less securely identified
by using only one observation at the time.
We also find that with the new pipeline we can robustly identify clusters with a central AGN that would
otherwise have been missed, and we can flag pairs of AGNs close on the sky that might have been mistaken
for a cluster.}
{}


\titlerunning{The new XXL pipeline}

\maketitle

\section{Introduction}
\label{sec:intro}

The XXL survey is a large-scale survey of the X-ray
sky carried out with the XMM-Newton satellite and designed both
to derive competitive constraints on cosmological parameters,
especially for the Dark Energy equation
of state, see \citet{pierre11}, and to provide a rich legacy data set.
The survey has observed two $\approx 25 \desqg$ patches of sky with good coverage
across multiple wave bands.
The rationale for such a survey is thoroughly explained in \citet{pierre16},
hereafter XXL Paper I, to which we refer for details; here we recall two
facts from XXL paper I relevant to the present paper.
First, the survey flux sensitivity in $[0.5-2]~\kev$, the band most relevant to
cluster studies and the one which we use to test
our pipeline, is $6\times 10^{-15}\funits$ ($90\%$ completeness limit)
for point sources \citep[][XXL Paper XXVII]{chiappetti18}.
Second, the survey layout is made up of XMM-Newton observations
(hereafter referred to as `pointings') separated both in right ascension (RA)
and declination (DEC) by $20\arcmin$ so as to have good overlap among them
given that the XMM field of view (FoV) is $\approx 30\arcmin$ in diameter.
This tiling ensures good sensitivity over the whole survey footprint,
one of the strengths of XXL.

In the context of such a survey it is imperative to have a dedicated pipeline
for the identification of extended sources, as the default XMM-Newton software
developed for such purposes is not really optimised for the relatively faint
clusters which XXL mostly observed with shallow exposures.
The need for such a pipeline was recognised early on in the course of
the XMM-LSS survey \citep{pierre04}, the forerunner to XXL, and
\citet{pacaud06} (hereafter P06) introduce a dedicated survey pipeline, called
\xamin~in P06 and hereafter referred to as \xaminP.
P06 provide a thorough description of \xaminP~and of the
rationale behind its development, extensively test it via simulations, and use it
to define the XMM-LSS selection function.
Since its introduction \xaminP~has been successfully used by the XMM-LSS project
to assemble their cluster sample \citep{pierre06, pacaud07, willis13, clerc14}
and the survey source catalogues, both point-like and extended
\citep{pierre07, chiappetti13}.
In the XXL project \xaminP~has been so far used to define the brightest 100
cluster sample \citep[][XXL Paper II]{pacaud16} and the 1000 brightest point source
sample \citep[][XXL paper VI]{foto16}, as well as the catalogue of 365 clusters
\citep[][XXL Paper XX]{adami18}, and the newer source catalogue (XXL Paper XXVII).
The pipeline was also used by the X-CLASS project \citep{clerc12} to perform a
complete reprocessing of the whole XMM-Newton archive with the aim of building
a cosmologically useful sample of serendipitous X-ray galaxy clusters.

The major shortcoming of the otherwise very satisfying \xaminP~is that it works
on each XMM-Newton pointing separately; it is  possible for a source not
to be detected, or to be badly measured, simply because it lies at a large
off-axis angle in that observation, where the sensitivity of the X-ray telescopes
is sharply degraded.
It is therefore necessary to address this weakness, as it would prevent us from
making optimal use of the carefully designed XXL survey layout.
As a concrete example, let us consider the case (relevant for the survey: we recall that
the centres of the pointings are spaced by $20\arcmin$
on the sky) of a source lying halfway between two such pointings.
This source would be at an off-axis angle of $10\arcmin$ from the centres of
both pointings, where the XMM-Newton telescopes sensitivity is degraded by about
$\approx 50\%$; therefore, with two observations the number of counts collected
would be about the same as the counts collected by a single observation on-axis
(for the same exposure time) but distributed over the two pointings.
By analysing each of these two pointings separately, as done by \xaminP, there
is the risk that the source will not be detected at all or, even if it is
detected, that there will be too few counts to securely classify it.
However, by combining both observations (thus using all the counts from all the
pointings at the same time) the source may be clearly detected and characterised.
We have therefore completely rewritten the pipeline to allow it to optimally use all
X-ray information by using all the available pointings at the same time; as this new
pipeline is largely new we will refer to it as \xaminF.

The paper is organised as follows: Section \ref{sec:xamin} describes the
\xaminF~pipeline; Section \ref{sec:sources} describes how to identify different
types of sources; Section \ref{sec:sim} describes the suite of simulations we use
to test \xaminF; Section \ref{sec:ov} describes our results; Section \ref{sec:p1}
describes a new selection for an almost pure sample of point sources (the \emph{P1}
selection) and summarises all selection criteria we introduce for the
different sources considered in the paper;
and Section \ref{sec:conclusion} presents our conclusions.
In the following we  use the terms `point source' and `AGN' interchangeably
as virtually all extra galactic X-ray point sources are indeed AGNs.
For the same reason we  use the terms `extended source' and `cluster'
interchangeably.

\section{The \xaminF~pipeline: general description}
\label{sec:xamin}

\citet{pacaud06} provide a thorough introduction to \xaminP; however, we  repeat here
several points for the sake of being self-contained.

\subsection{Event lists}

Calibrated event lists are created from raw observation data files (ODFs) using
SAS tasks
{\texttt{emchain}} and {\texttt{epchain}}, and are
then filtered for solar soft proton flares.
Photon flares are filtered using the light curves of high-energy events; the band
used for filtering differs for each
EPIC detector: $10-12~\kev$ for the MOS1 and MOS2 detectors \citep{turner01}
and $12-14~\kev$ for the pn detector \citep{struder01}.
Histograms of these light curves are created, binned by $104~\mathrm{s,}$ and fitted to a Poisson
law of mean $\lambda$, and intervals where emission exceeds
$\lambda + 3\sqrt{\lambda}$ are discarded; this method is described by \citet{pratt02}.
Soft proton cleaned lists are then used to produce images of
$2.5\arcsec/\mathrm{pixel}$ to  correctly sample  the XMM-Newton PSF
($\approx 6\arcsec$ on-axis), using the 
SAS task {\texttt{evselect}}

One image for each EPIC detector (MOS1, MOS2 and pn) is then created for each energy
band of interest: $[0.3-0.5], [0.5-2], [2-10]~\kev$; all these steps are
unchanged from \xaminP~and more details are given in P06.
In the following we consider only $[0.5-2]~\kev$ images as this is the most
important band for cluster detections and characterisation, especially for the
faint clusters typically observed by XXL.

\subsection{Tile creation}

Next, mosaicked images of the XXL sky, hereafter referred to as {`tiles'} 
(the term `mosaic' is  reserved for combined images, tiles, or single
pointings of more than one EPIC instrument), are created, one per EPIC instrument.
These tiles, which are  \atile~images  pixelized at $2.5\arcsec$, are obtained by
re-projecting the event lists pertaining to each tile to a common frame in the sky
using SAS task {\texttt{attcalc}}.
Tiles are spaced by $60\arcmin$ each in RA and DEC; the $68\arcmin$ length
allows for a $4\arcmin$ overlap across tiles.  

As the XXL coverage is rather dense (we recall that $30\arcmin$ diameter pointings
are separated by $20\arcmin$ in RA and DEC),
on average $\approx 20-25$ pointings are included in one $\desqg$,
most of which only partially overlap with the tile; conversely, each pointing may
in general fall across tile borders and therefore can be used in many tiles.
For each \atile~region of sky covered by the survey we now have three tiles (one
for each instrument); these three tiles are co-added to have a single
MOS1$+$MOS2$+$pn \atile~tile which is then wavelet smoothed as the first stage of
source detection.
Combined exposure maps and detector masks for each tile are also created at this
stage.
This step is new in \xaminF.

\subsection{Preliminary source detection}

We are now ready to perform preliminary source detection using
the mosaicked MOS1$+$MOS2$+$pn \atile~tiles to take advantage of the many
overlapping pointings in a tile.

We use the mixed approach introduced by \citet{valtchanov01} (hereafter V01)
and used in P06; it consists of first filtering the input X-ray
image and then performing source detection on the smoothed image, taking
advantage of the many well-developed source detection procedures
developed for optical images.
The validity of this approach is demonstrated via extensive simulations in
V01, who show that it gives the best results compared to the other
approaches they test for detecting and characterising
both point-like and extended objects when used on XMM-Newton images.
Filtering an image through wavelets is a popular choice and many techniques
have been introduced. \citet{starck98} (hereafter SP98) show via simulations that
the best filtering method for images containing Poisson noise with few photons
like X-ray images is the method based on the auto-convolutions of the histogram of
the wavelet function.
SP98 show in particular that the method is effective at recovering
extended sources with only few photons which is the case of XXL clusters.
They also note that one strong point of the method is that it does not
need a background model,  and they show how an input cluster can be successfully
retrieved with different background levels.
We refer to SP98 for the relevant formulas and we only briefly
recall the main features of the method in Section \ref{subsec:mr1}.

\citet{mr1} describe their implementation of the method in the
\texttt{MR/1} software; we chose to develop our own IDL implementation of
\texttt{MR/1} (the rest of \xaminF~is written in Python),
based on the latest implementation kindly provided by Jean-Luc Starck,
to make it easier to perform two additional steps, not executed in \xaminP, before smoothing;
these steps are detailed below.

\subsubsection{File preparation}
\label{subsec:file}

As a first step a model of particle background, obtained by simulating very long
XMM-Newton exposures with closed filter wheel, is subtracted from the tiles.
This is necessary because the background component is unvignetted (the probability
of a particle being mistakenly detected as a photon is independent of off-axis
angle) so there is an excess of photons due to particle background in regions of
overlap with respect to the centre of the pointings.
This excess, if not corrected, may bias source detection; the problem is absent
when using images of single pointings as in \xaminP.
To model the particle background we use the same procedure used to add
particle background on simulated images, but using an exposure time of
$1~\mathrm{Ms}$ to ensure a good sampling of all pixels;  a nominal XXL
$10~\mathrm{ks}$ exposure time would have left many pixels empty.
We then re-scale the pixel values to the nominal XXL exposure time $10~\mathrm{ks}$.

The second step consists in correcting for the different exposure time across a
tile by dividing it by a tiled, combined exposure map; this step was not performed
by \xaminP, but we found that it improves overall performance.
The combined exposure map is obtained by adding together the tiled exposure maps
of the three EPIC detectors.
We  compensate for the fact that the effective area of the pn
detector is $\approx 3.1$ times the effective area of the MOS detectors in the energy
range of interest ($[0.5-2]~\kev$) by multiplying the pn tiled exposure map by
$3.1$ before adding it to the MOS ones.
This last step is useful because it returns a lower estimate of the count rate of a
candidate source than  would be obtained by simply adding the three exposure
maps, since the exposure time is longer.
This lowers the risk that the fit will go wrong because \sex~returns  an
initial count rate estimate that is too high, a problem that sometimes occurs. 
Figure \ref{fig:particlebgd},  left panel, shows an example of particle
background model for a simulated \atile~tile reproducing the real XXL tiling;
we note  the strong spatial variation due to the fact that this background component
is unvignetted so it is much more prominent in regions of overlap.
In the right panel of Figure \ref{fig:particlebgd}  the tiled combined
(MOS1$+$MOS2$+3.1\times$pn) exposure map is shown of the same simulated region of sky
assuming the nominal exposure time $10\mathrm{ks}$ for each instrument.
The map is given in MOS units: we correct for the larger effective area of the
pn detector by multiplying it by $3.1$ so at the centre of each pointing the
total exposure time is $\approx 50\mathrm{ks}$.
This is roughly the time needed to reach the nominal XXL sensitivity in the
$[0.5-2]~\kev$ band with a single MOS detector.

\subsubsection{Wavelet smoothing}
\label{subsec:mr1}

\begin{figure*}
        \centering
        \includegraphics[width=\linewidth]{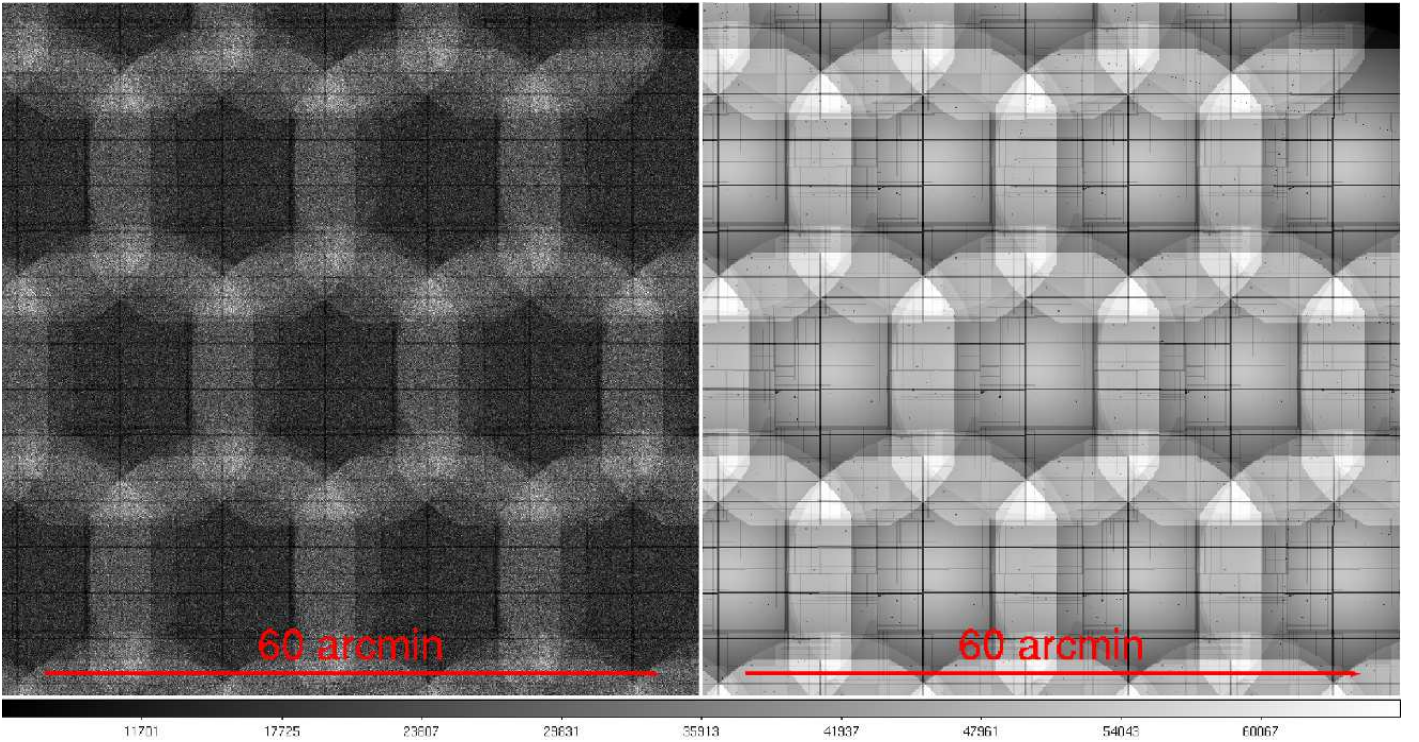}
        \caption{Left: Example of particle background model subtracted from a \atile~tile.
        Right: Tiled combined exposure map (MOS1$+$MOS2$+3.1\times$pn)  of the same region.
        A nominal XMM-LSS exposure time $10~\mathrm{ks}$ for each instrument is assumed.
        The map is in MOS units: the pn exposure map is corrected for the larger pn
        effective area by multiplying it by $3.1$ (see text) so the combined exposure map
        has an exposure time $\approx 50~\mathrm{ks}$ at the pointing centres.
        The exposure time is even larger ($\approx 60~\mathrm{ks}$)
        in areas where three pointings overlap.
        }
        \label{fig:particlebgd}
\end{figure*}

After performing the first two steps detailed above we are
ready to wavelet smooth the tile.
The wavelet smoothing procedure is applied to $(I - P) / \mathrm{Expo}$, where $I$
is the input MOS1$+$MOS2$+$pn tile, $P$ is the particle background model image
(see the left panel of Figure \ref{fig:particlebgd}), and $\mathrm{Expo}$ is the
combined tile exposure map (see the right panel of Figure \ref{fig:particlebgd}).

The \texttt{MR/1} algorithm computes a smoothed image from an input image
as the sum of a predefined number of scale dependent wavelet coefficients; each
coefficient is computed from the input image using a chosen wavelet function
$\psi(x,y)$, in our case a B3-spline.
A wavelet coefficient $w_j(x,y)$ at position ($x,y$) and scale $j$ (the total
number of scales is fixed) carries information about whether or not there is signal
in the image at that position and scale; only coefficients that contain signal,
according to the criteria detailed below, are included in the final smoothed image.
Many different choices of $w_j(x,y)$ have been considered in the literature;
SP98 review several of them and show that, for the Poisson regime
typical of X-ray images, a good choice of $w_j(x,y)$ is the one
introduced by \citet{slezak93} and \citet{bury95} in the context of
galaxy clustering; its expression is given by Equation $6$ of SP98 which
we reproduce here:

\begin{equation}
\label{eq:w}
w_j(x,y)=\sum_{k \in K} n_k\psi\Bigg (\frac{x_k - x}{2^j},\frac{y_k - y}{2^j}\Bigg).
\end{equation}
In Equation \ref{eq:w} $K$ is the support of the wavelet function $\psi$,
i.e. the box where $\psi \neq 0$, and $n_k$ is the number of photons at position
$(x_k,y_k)$; we note that $\psi$ is dilated by a factor $2^j$ for scale $j$ and
is centred at the coefficient position ($x$,$y$).

It is easy to understand from Equation \ref{eq:w} why this choice is
appropriate for the Poisson regime: $n_k$ can take any value, including $0$,
and no assumption, e.g. of Gaussianity, needs to be made.
If the coefficient $w_j(x,y)$ is due to noise
it can be considered  the sum of $n$ independent random variables, each
corresponding to one count and having a distribution given by the histogram
$H_1$ of the wavelet function $\psi$.
The distribution of the sum of $n$ such independent variables is given by $n$
auto-convolutions of $H_1$:

\begin{equation}
\label{eq:hn}
H_n=H_1\otimes H_1\otimes\ldots H_1
\end{equation}
So a simple and rigorous way of assessing the significance of $w_j(x,y)$
is to compare it to a threshold expressed as a pre-defined multiple
(denoted by $\sigma$) of the standard deviation of $H_n$.
Only coefficients found to be significant according to this criterion are
included in the smoothed image.
$H_n$ converges to a Gaussian distribution in the limit of large $n$.

Equation \ref{eq:w} shows that each scale is twice
as large as the previous one (each scale $j$ corresponds to $2^{j-1}$ pixels);
as in P06 we consider eight scales (scales $2-9$; scale $1$ is not used) with dimensions varying from $2$ to $256$ pixels.
We impose a set of eight thresholds with values $\sigma=3,3,3,3,4,4,3,3$ for the eight
scales  used.  We choose a $3\sigma$ threshold for most scales; however, for scales
$6$ and $7$ (which correspond to $32-64~\mathrm{pixels}=80\arcsec-160\arcsec$, roughly the
scale of the largest clusters observed by XXL)
we demand a higher threshold for significance.

We found that doing so improves overall performance and allows us to reduce the
number of sources mistakenly identified as clusters because their photons are spread
out to a large area by the strong XMM-Newton PSF distortion at large off-axis
angles.

Our choice of thresholds differs somewhat from that of  P06: they define a $P$ value such
that, if the probability for a wavelet coefficient $w_j(x,y)$ being due to noise
(computed from the cumulative distribution of $H_n$) is $<P$, the coefficient is
considered significant; P06 choose $P=0.001$ for all scales, which, in the limit of
large $n$, corresponds to $\sigma~\approx 3.09$.

It is important to note that the \texttt{MR/1} wavelet filtering
removes noise already in the wavelet filtering stage thanks to the thresholding
scheme adopted; the result is a smoothed image in which most of the noise is
already almost completely removed and the background has been smoothed;
this justifies the use of \sex~in source detection described in
Section \ref{subsec:sex}

\subsubsection{Identifying sources on the smoothed image using \sex}
\label{subsec:sex}

To identify candidate sources in the smoothed image we use
\sex~\citep{bertin96};
the suitability of this package is demonstrated by V01 who
use it in conjunction with \texttt{MR/1} to successfully recover point and
extended sources in simulated XMM-Newton images.
V01 present detailed statistics of missed or false source detections
for their simulations in which they show that
the \texttt{MR/1}$+$\sex~combination is the best overall of the many approaches
they study; in particular, they show that it works better than all the other
methods in recovering high-$z$ clusters, an important concern for XXL.

 \citet{valtchanov01} also present  a detailed discussion of the many possible choices of
\sex~parameters, identifying the choices that give the best results
and which are adopted in P06.
\sex~parameters are unchanged from P06 except  that we use a larger box
(512 pixels instead of 64) to estimate the background.
V01 point out that the choice of a good background box size is a tricky one,
implying a trade-off between bad photometry (small box) and the risk of missing
faint sources (large box); they advocate boxes of $32-64$ pixels, and P06
adopt $64$ pixels.
We found however that choosing a  background box as large as $512$ pixels helps
to reduce the number of false cluster detections when using tiles and does not
provoke the loss of faint sources, but instead improves things a bit.
In our simulations we find  for $25\deg^2$ that using a $512$ pixel box
leads to recovering $\approx 8200$ genuine point sources, whereas using $64$
pixels leads to recovering $\approx 7800$ genuine ones.
We conclude that using a  $512$ pixel background box does not have an adverse
effect on pipeline performance regarding point sources and is beneficial
regarding extended ones.
See Section \ref{sec:cash} for the criteria  to decide whether a
detection is genuine or due to background, and Section
\ref{sec:sim} for details on simulations.

Table \ref{tab:param} reproduces Table 1 of P06 and reports
the values of the parameters used for the detection stage; most parameters are
unchanged from P06; the cases where they
are not are explicitly noted.

\begin{table*}
\begin{center}
\caption{Relevant parameters of the XXL pipeline detection stage.
\label{tab:param}}
\begin{tabular}{lll}
\hline
Parameter & value & comment\\
\hline

\multicolumn{3}{l} {\bf Event selection:}       \\
MOS event flag selection                        &       \#XMMEA\_EM & Same as P06\\
pn event flag selection                         &       (FLAG \& 0x2fb002c)==0 & Same as P06\\
MOS patterns                                    &       [0:12] & Same as P06\\
pn patterns                                     &       [0:4] & Same as P06\\
\hline

\multicolumn{3}{l} {\bf Image:}                 \\
Type                                            &       sky & Same as P06\\
Configuration                                   &       co-addition of EPIC detectors & Same as P06\\
Pixel size                                      &       2.5\arcsec & Same as P06\\
\hline

\multicolumn{3}{l} {\bf MR/1:}                  \\
Wavelet type                                    &       B3-spline & Same as P06\\
Transform algorithm                             &       ``\`a trou" & Same as P06\\
Thresholds($\sigma$)                            &       $3,3,3,3,4,4,3,3$ & in P06: $P<10^{-3}$ \\
Lowest significant scale                        &       $2$ pix. & Same as P06\\
Highest significant scale                       &       $256$ pix. & Same as P06\\
\hline

\multicolumn{3}{l} {\bf {\texttt{SExtractor}:}} \\
Background cell side                            &       512 pix. & In P06: 64 pixels\\
Background median filtering                     &       4 cells & Same as P06\\
Detection threshold                             &       6$\sigma$ & Same as P06\\
Detection minimum area                          &       12 pix. & Same as P06\\
De-blending sub-thresholds                      &       64 & Same as P06\\
De-blend min. contrast                          &       0.003 & Same as P06\\

\hline
\end{tabular}
\end{center}
\end{table*}

\subsection{Likelihood fit: models and parameters}
\label{sec:fit}

After \sex~has found the list of candidate sources for each \atile~tile, each candidate
source is characterised by performing a maximum likelihood fit based on the $C$-statistic
\citep{cash79}, appropriate for the Poisson regime, using raw photon images; all images
that cover the candidate are used, whereas with \xaminP~images were used one by one.

It is essential to understand that for the likelihood fit
tiles {cannot} be used because when fitting it is necessary to correctly account
for the strong PSF distortion introduced by the XMM-Newton telescopes, which depends on
off-axis and position angles and so is different for each individual pointing, as 
detailed below.
Each source is fit to the following:

\begin{enumerate}

\item a PSF model (\textsc{pnt} fit);

\item a $\beta$ model \citep{cavaliere76} (\textsc{ext} fit);

\item a $\beta$ model superposed to a central PSF (\textsc{epn} fit);

\item two PSFs (\textsc{dbl} fit);

\end{enumerate}
the $\beta$ model is also convolved with the XMM-Newton PSF.

All PSF models used in all fits are computed at the
source position on each different pointing; since these positions are
in general different from pointing to pointing, a different PSF model must be used for
each different pointing, preventing the use of a single tile where photons from different
pointings, differently affected by the PSF distortion, are grouped together.

In all cases a local background is estimated by subtracting the total number
of photons expected from the model from the total number of photons
in the fit region.
A model in which all photons are assumed to be due to background is also
considered; we refer to it as the \textsc{bkg} model and it is used
in assessing the significance of each fit in connection with Equation
\ref{eq:detstat}.

The \textsc{pnt} and the \textsc{ext} fits are present in \xaminP;
the \textsc{epn} and the \textsc{dbl} fits are new to \xaminF~and will be described in
more detail in Subsection \ref{sec:epnfit}.

The $\beta$ profile is described by

\begin{equation}
\label{eq:beta}
S_\mathrm{X}(r)\propto\bigg[1+\bigg(\frac{r}{\textsc{ext}}\bigg)^2\bigg]^{-3\beta+1/2},
\end{equation}
where the \emph{core radius} \textsc{ext} is measured in $\arcsec$ and $\beta=2/3$.
Different values of $\beta$ may be specified at the start of the fit, but its value
is then kept fixed as, in general, XXL clusters have too few counts to robustly
constrain it. 

Fit parameters always include count rates $CR_\mathrm{mos}$ and
$CR_\mathrm{pn}$; MOS1 and MOS2 are assumed to be identical
and only one count rate for them is introduced. In the \textsc{ext} fit
the core radius \textsc{ext} is also a fit parameter; in the \textsc{epn}
and \textsc{dbl} additional parameters are introduced as explained in
Subsection \ref{sec:epnfit}.
A total count rate $CR$ can be estimated from
$CR_\mathrm{mos}$ and $CR_\mathrm{pn}$ as

\begin{equation}
\label{eq:cr1}
CR = 2\times CR_\mathrm{mos} + CR_\mathrm{pn},
\end{equation}
and a total count rate $CR_\mathrm{MOSNORM}$ normalised to the
MOS effective area can also be estimated as

\begin{equation}
\label{eq:cr2}
CR_\mathrm{MOSNORM}=2\times CR_\mathrm{mos} + \frac{CR_\mathrm{pn}}{3.1}.
\end{equation}

Equation \ref{eq:cr2} allows us to compensate for the difference in sensitivity
between the MOS and pn detectors; in the following when quoting values for derived
count rates we  use Equation \ref{eq:cr1}.
All fit parameters are forced to be the same across all pointings: there is only one
$CR_\mathrm{mos}$, one $CR_\mathrm{pn}$, one \textsc{ext}, and so on.
In principle the source position may be fitted in all fits; however, 
unlike P06 where it was fitted in the \textsc{ext} fit and kept fixed in the
\textsc{pnt} fit, we do not fit for it in any case but we always keep it fixed at
the value found by \sex.

\subsection{Likelihood fit: the $C$- and $E$-statistics}
\label{sec:cash}

The validity of each model in the Poisson regime can be estimated using the Cash $C$-statistic \citep{cash79}

\begin{equation}
\label{eq:cash}
C=2\sum_{i=1}^{N_{pix}} m_i - y_i\ln{m_i},
\end{equation}
where $y_i$ is the number of observed photons and $m_i$ is the
number of photons expected from the model in pixel $i$.
Introducing
$N_\mathrm{pix}$ (the number of pixels used in the fit),
$N_\mathrm{data}\equiv\sum_{i=1}^{N_\mathrm{pix}} y_i$ (the total number of observed 
photons used in the fit), and $N_\mathrm{mod}\equiv\sum_{i=1}^{N_\mathrm{pix}} m_i$
(the overall model normalisation), we can express the value of each model pixel
$m_i$ as $m_i=N_\mathrm{mod}\times d_i$ with $d_i\equiv\frac{m_i}{N_\mathrm{mod}}$,
and we can rewrite Equation \ref{eq:cash} as

\begin{equation}
\label{eq:cash2}
C=2 \left(N_\mathrm{mod} - N_\mathrm{data}\ln{N_\mathrm{mod}} \right) - 2\sum_{i=1}^{N_{pix}} y_i\ln{d_i},
\end{equation}
which has the advantage of explicitly factoring out $N_\mathrm{data}$ and
$N_\mathrm{mod}$.
Minimising Equation \ref{eq:cash2} with respect to $N_\mathrm{mod}$ yields
$N_\mathrm{mod}=N_\mathrm{data}$ and we choose to fix
$N_\mathrm{mod}=N_\mathrm{data}$ and use the simplified $E$-statistic

\begin{equation}
\label{eq:e}
E = - 2\sum_{i=1}^{N_{pix}} \left( y_i\ln{d_i}\right),
\end{equation}
to assess the validity of each model (convolved with the XMM-Newton PSF) used in
the fit.

We use the $E$-statistic because it allows us to have one fewer
parameter in the fit, since the model normalisation $N_\mathrm{mod}$ is now fixed at
$N_\mathrm{data}$; since XXL observations do not
have many photons it is important to reduce the number of parameters when possible.

The $E$-statistic is equivalent to the $C$-statistic for
parameter estimation, but not for uncertainty estimation; as we are not interested in
uncertainty estimation (see Subsection \ref{sec:error}) this is not a serious
problem.
It should be noted that the $C$- and $E$-statistics are not really likelihood
functions but are related to the likelihood function $L$ by
$C=-2\log L + \mathrm{const}$ (and similarly for $E$).
It is instead more appropriate to think of them as the
Poisson distribution equivalent of the $\chi^2$ statistic appropriate to the
Gaussian distribution (recall that $\chi^2=-2\log L + \mathrm{const}$ as well)
and, like $\chi^2$, $C$ and $E$ reach a minimum at the optimum.

The significance of a detection is assessed by evaluating the increase
in $E$ between its best fit value $E_\mathrm{BF}$ and a model containing
only background (the \textsc{bkg} model introduced above); this allows us to
define the {detection statistic}

\begin{equation}
\label{eq:detstat}
\textsc{det\_stat} = 2N_\mathrm{data}\ln(N_\mathrm{pix})-E_\mathrm{BF},
\end{equation}
used by P06 to show via simulations that a good criterion for discriminating
real detections from chance background fluctuations is given by

\begin{equation}
\label{eq:sd}
\textsc{det\_stat} > 15;
\end{equation}
these sources were referred to in P06 and subsequent papers
as `point sources'.

This criterion is also valid for \xaminF~and we use Equation \ref{eq:sd}
to discriminate between real detections and chance background fluctuations.
Applying the cut of course leads to several faint real sources to be found by
\sex~and then discarded.
In our simulations we find that for $25\deg^2$ this happens for
$\approx 1400$ input sources out of every $\approx 10000$ found by
\sex.
In Section \ref{sec:p1} we give more information about the selection of
point sources.

All these considerations apply to each of the four fits we perform, and
we derive one value of \textsc{det\_stat} for each of them; we must then
compare different fits to one another to decide which best describes the
source.
To assess the significance of an \textsc{ext} fit over a \textsc{pnt} fit,
P06 introduce an {extent statistic} defined as

\begin{equation}
\label{eq:extstat}
\textsc{ext\_stat} = (E_\mathrm{BF})_\mathrm{PNT} - (E_\mathrm{BF})_\mathrm{EXT},
\end{equation}
which continues to be used in \xaminF; $\textsc{ext\_stat}$ plays a crucial
role in cluster identification
The issue of selecting a best fit model among the four models we consider
is very important and is described in more detail in Section \ref{sec:sources},
after introducing the \textsc{epn} and \textsc{dbl} fits.

\subsection{Likelihood fit: the \textsc{epn} and \textsc{dbl} fits}
\label{sec:epnfit}

Whereas the \textsc{ext} and \textsc{pnt} fits are unchanged from
\xaminP, apart from the above-mentioned improvements in wavelet smoothing,
and other improvements described in Subsection \ref{sec:other},
the \textsc{epn} and \textsc{dbl} fits are new and are described here.

The \textsc{epn} is introduced to allow the recovery of clusters with
a strong contamination by a central AGN; in Subsection \ref{sec:ac}
we show how, in this case, a real cluster can be missed by the
\textsc{ext} fit, which means that it is necessary to introduce a more sophisticated 
fit to recover it.
The fit can also be used to flag clusters which, though being identified as such
by the \textsc{ext} fit, are nevertheless contaminated by central AGN; these  clusters may be interesting in themselves.

In the \textsc{epn} fit we fit the source to a superposition of a $\beta$ profile
(always convolved with the XMM-Newton PSF) and the PSF itself, placed at the
cluster centre to model the effect of the central contaminating AGN.
Again, the value of $\beta$ can be specified in advance and we choose
$\beta=2/3$, but it is then kept fixed during the fit.
The fit parameters are
$CR_\mathrm{mos}$ and $CR_\mathrm{pn}$, 
the cluster core radius, called \textsc{epn\_ext} to distinguish
it from the core radius $\textsc{ext}$ computed by the $\textsc{ext}$ fit, and the
relative count rate between cluster and AGN, indicated by \textsc{epn\_ratio},
which is the same for the MOS and pn detectors; we do not fit for any offset
between the AGN and the cluster as in general we do not have enough counts to
constrain it.
The count rates for the cluster and the AGN can be computed from
the total count rate $CR$ given by Equation \ref{eq:cr1}
and \textsc{epn\_ratio} as

\begin{eqnarray}
\label{eq:epncr}
\nonumber
CR_{Cluster} &=&CR / (1 + \textsc{epn\_ratio}) \\
CR_{AGN} &=&\textsc{epn\_ratio}\times CR/(1+\textsc{epn\_ratio}). \\
\nonumber
\end{eqnarray}

The \textsc{dbl} fit is introduced to account for the frequent case
where two or more AGNs are close in the sky and their combined X-ray emission
can be mistakenly classified as a single extended emission in the \textsc{ext};
the fit is used in conjunction with the cluster classification criteria in
Subsection \ref{sec:clustersel} to flag such cases since we obviously do
not want to include them in any subsequent cluster catalogue.
The \textsc{dbl} fit is performed in a  manner similar to that of  the \textsc{epn} fit:
the source is fit to a superposition of two XMM-Newton PSFs; the fit parameters are
$CR_\mathrm{mos}$ and $CR_\mathrm{pn}$,
the relative count rate between the PSFs \textsc{dbl\_ratio},
which again is the same for the MOS and pn detectors, and
the separation between the PSFs \textsc{dbl\_sep}.
We fit only the separation \textsc{dbl\_sep}; the midpoint of the line joining the
PSFs is always the source \sex~position, which we keep fixed.
The position angle between the PSFs is computed in advance by first smoothing the
MOS1$+$MOS2$+$pn image of each pointing using a
Gaussian filter with $\sigma=2~\mathrm{pixels}$ and then computing its second
moments.
We note that in doing so we estimate a different position angle for each pointing.
Unlike \textsc{dbl\_sep}, the position angle is then kept fixed at the value
computed from its second moments and it is not a fit parameter because in general
we do not have enough counts to constrain it.
The relative count rate of the two point sources can be computed from an equation
similar to Equation \ref{eq:epncr}.

\subsection{Parameter uncertainties}
\label{sec:error}

We have not said anything about the uncertainty in the recovered parameters.
The reason is that \xaminF~is primarily a {detection pipeline} to identify
sources as real and securely classify them according to their type (point or
extended); it is {not} meant primarily to derive accurate values of
source parameters and their associated uncertainties
for the cluster candidates.
More accurate parameter estimation is carried out in successive steps
by other means such as growth curve analysis for count rates, as done by e.g. \citet{clerc12}, and in the XXL project, in XXL paper II, and by
\citet[][XXL paper III]{giles16}.
It is still possible, however, to use the pipeline in either version to derive
uncertainties using simulations: \citet{pierre07} and \citet{chiappetti13} derive
estimates of the positional accuracy of the sources in the first and second version
of the XMM source catalogue, respectively, as a function of the count rate derived
by the pipeline and off-axis angle, by running \xaminP~on simulated XMM-Newton
images.
Flux uncertainties, on the other hand, are currently estimated,
as explained in XXL Paper XXVII, by calculating the Poisson error
on gross photons according to the formula of \citet{gehrels86}.
Gross photons, in turn, are reconstructed by adding net photons and background
photons in the fitting region, as computed by \xaminF.

\subsection{Choosing the fitting region}
\label{sec:box}

\sex~returns an estimate of the source extent via the ellipse parameters
$\textsc{CEA\_A}$ and $\textsc{CEA\_B}$, the semi-major and semi-minor
axes, in arcseconds, of the ellipse that best describes the source.
We choose an initial fitting region as $3$ times the mean 
semi-axis $(\textsc{CEA\_A}+\textsc{CEA\_B})/2$, and if there are
no photons in it we take a region three times as large; we always impose
the constraint that the region must be $>35\arcsec$ and $<200\arcsec$.
Within this region we flag pixels belonging to different sources according
to the \sex~pixel segmentation mask; these pixels are excluded from the fit
which is then not affected by neighbouring sources.

\subsection{Other improvements}
\label{sec:other}

Other important improvements are the following:

\begin{enumerate}

\item
\xaminF~uses the latest XMM-Newton PSF model, described in \citet{read11}, whereas
\xaminP~used the older `Medium' model composed of a set of images, the same for each of
the three XMM-Newton telescopes, which did not take into account the strong azimuthal
dependence of the PSF shape.

\item
\xaminP~was not optimised for detection of bright point sources and
sometimes it missed them; \xaminF~corrects this and is able to detect
bright point sources more reliably.
It should be noted that since cluster identification is based on the
computation of the extent statistic (Equation \ref{eq:extstat}), which
requires the results of the \textsc{pnt} fit, a failure of the \textsc{pnt} fit
may cause a failure to recognise a candidate source as a cluster; as a consequence,
several bright clusters in the XXL images were missed by \xaminP~which
are now detected by \xaminF.

\end{enumerate}

\section{Identifying different types of sources}
\label{sec:sources}

\xaminF~performs four fits on each candidate source
and computes a detection statistic \textsc{ext\_det}  for each of them (see Sect.
\ref{sec:fit}). We  indicate these statistics
by \textsc{pnt\_det\_stat}, \textsc{ext\_det\_stat}, \textsc{epn\_det\_stat},
\textsc{dbl\_det\_stat}; all the statistics are dimensionless.

Intuitively a source will be flagged as point, extended, extended$+$point, or
double depending on which of its detection statistics mentioned above is highest
and by how much.
To do this rigorously we must first introduce the appropriate
statistics, the analogue of \textsc{ext\_like} in the \textsc{ext} fit, and then
we must derive quantitative criteria of source classification based upon these
statistics by using simulations.
The relevant statistics are introduced in Subsections \ref{sec:clustersel} and
\ref{sec:acsel}.

The criteria for cluster selection, the \emph{C1} and \emph{C2} selections introduced 
in P06, are reported in Subsection \ref{sec:clustersel} and are shown to be still valid for \xaminF\ (see
Sect.
\ref{sec:clusters}).
The criteria for recovering clusters contaminated by a central AGN are defined and 
tested via simulations in Subsection \ref{sec:ac}.
The criteria for flagging double sources that may be misidentified as extended ones are
introduced and tested via simulations in Subsection \ref{sec:doubles}.
The criteria for selecting an almost pure sample of point sources (the \emph{P1}
selection) are introduced and tested via simulations in Section \ref{sec:p1}.
Finally, in Table \ref{tab:sources} we summarise all the criteria we introduce.
These criteria are usually defined by several conditions;
if more than one condition is specified all conditions must be used
unless explicitly stated otherwise.

\subsection{Selecting clusters}
\label{sec:clustersel}

To select clusters, as explained in P06 and restated in subsection \ref{sec:fit}, we
introduce an extent statistic (\textsc{ext\_stat}), defined in Equation
\ref{eq:extstat} as the difference between the best fit detection statistics of
the \textsc{pnt} and \textsc{ext} fits.
In principle \textsc{ext\_stat} can have either sign, and we expect that for a point
source it should be negative; if \xaminF~finds a negative value of \textsc{ext\_stat}
it forces it to $0$, so we expect a point source to have $\textsc{ext\_stat}=0$ in
the \textsc{ext} fit and in most cases this is true.
In certain cases, however, a point source will have $\textsc{ext\_stat}>0$ and we must
establish a threshold to decide whether a certain value of \textsc{ext\_stat} is
high enough for a source to be considered extended; this can only be done by
simulations.
P06 show via simulations that a threshold $\textsc{ext\_stat}>33$ allows us
to robustly distinguish between extended and point sources, and to introduce the
\emph{C1} selection:

\begin{eqnarray}
\label{eq:c1}
\nonumber
\textsc{ext}&>&5\arcsec, \\
\nonumber
\textsc{ext\_stat}&>&33, \\
\nonumber
\textsc{ext\_det\_stat}&>&32. \\
\end{eqnarray}
The C1 selection is shown in P06 to be almost pure; this continues to be true with
\xaminF.
P06 introduce a second selection, the \emph{C2} selection, composed of fainter
clusters and with a $\approx 50\%$ probability of contamination, defined as

\begin{eqnarray}
\label{eq:c2}
\nonumber
\textsc{ext}&>&5\arcsec, \\
\nonumber
\textsc{ext\_stat}&>&15. \\
\end{eqnarray}
We  show in Subsection \ref{sec:clusters} that the C1/C2 selections as defined in
Equations \ref{eq:c1} and \ref{eq:c2} are still appropriate for \xaminF.

\subsection{Recovering AGN contaminated clusters and flagging double sources}
\label{sec:acsel}

For the \textsc{dbl} fit we introduce, in the same vein, a {double statistic}
(\textsc{dbl\_stat}) defined as the difference between the best fit detection
statistics of the \textsc{pnt} and \textsc{dbl} fits (see Equation
\ref{eq:dblstat}):

\begin{equation}
\label{eq:dblstat}
\textsc{dbl\_stat} = (E_\mathrm{BF})_\mathrm{PNT} - (E_\mathrm{BF})_\mathrm{DBL}.
\end{equation}
This statistic  allows us to flag sources that may be initially identified as
extended (i.e. they pass the C1 or C2 criteria), but which are actually a double, as
is explained in Subsection \ref{sec:doubles}.

For the extended$+$point (\textsc{epn}) fit we need two statistics as we need to
quantify the likelihood of an \textsc{epn} fit with respect to a simple extended fit (\textsc{ext}),
and the likelihood of an \textsc{epn} fit with respect to a point fit (\textsc{pnt}).
The first statistic, which we  refer to as \textsc{epn\_stat\_ext}, is necessary
in order to classify a cluster contaminated by a central AGN as a cluster.
The second statistic, which we  refer to as \textsc{epn\_stat\_pnt}, is
necessary in order to distinguish an AGN contaminated cluster from a simple AGN; it may
happen that, for an AGN, an \textsc{epn} fit is better than an \textsc{ext} fit
(revealed by a high value of \textsc{epn\_stat\_ext}),  but not as good as a
\textsc{pnt} fit (revealed by a low or zero value of \textsc{epn\_stat\_pnt}).
To classify a source as a cluster contaminated by a central AGN
we require significant values of both statistics (defined in Subsection \ref{sec:ac} via simulations), whose equations are

\begin{eqnarray}
\label{eq:newstat}
\nonumber
\textsc{epn\_stat\_pnt}&=&(E_\mathrm{BF})_\mathrm{PNT} - (E_\mathrm{BF})_\mathrm{EPN}, \\
\nonumber
\textsc{epn\_stat\_ext}&=&(E_\mathrm{BF})_\mathrm{EXT} - (E_\mathrm{BF})_\mathrm{EPN}. \\
\end{eqnarray}

Table \ref{tab:par} summarises  the fit parameters and the statistics we
introduce.

\begin{table*}
\caption{Main fit parameters and statistics; all quantities in the table are
dimensionless except where explicitly noted.}
\label{tab:par}
\centering
\begin{tabular}{lllll}

\hline
Parameter           & Fit in which used & Comment & Units              & Present            \\
Name                &                   &         &                    & in \xaminP         \\
\hline

$CR_\mathrm{MOS}$        & \textsc{pnt},\textsc{ext},\textsc{epn},\textsc{dbl}      & MOS1,2 count rate & ($\mathrm{count/s}$) & Yes                \\  
$CR_\mathrm{pn}$         & \textsc{pnt},\textsc{ext},\textsc{epn},\textsc{dbl}      & pn count rate & ($\mathrm{count/s}$)     & Yes                \\
\textsc{ext}                 & \textsc{ext}                                             & core radius in \textsc{ext} fit & ($\arcsec$)        & Yes                \\
\textsc{epn\_ext}            & \textsc{epn}                                             & core radius in \textsc{epn} fit & ($\arcsec$)        & No                 \\  
\textsc{epn\_ratio}          & \textsc{epn}                                             & Count rate ratio between extended &         & No                 \\
                             &                                                          & and point source in \textsc{epn} fit &               &                    \\ 
\textsc{dbl\_ratio}          & \textsc{dbl}                                             & Count rate ratio between point    &         & No                 \\
                             &                                                          & sources in \textsc{dbl} fit          &               &                    \\ 
\textsc{dbl\_sep}            & \textsc{dbl}                                             & Separation between the two point  &        & No                 \\
                             &                                                          & sources in \textsc{dbl} fit & ($\arcsec$)            &                    \\     

\hline
Statistic           & Fit in which used & Comment            &   & Present in         \\
Name                &                   &                    &   & \xaminP            \\
\hline
\textsc{pnt\_det\_stat}  & \textsc{pnt}                 & Detection statistic, \textsc{pnt} fit    &           & Yes                \\
\textsc{ext\_det\_stat}  & \textsc{ext}                 & Detection statistic, \textsc{ext} fit    &           & Yes                \\
\textsc{epn\_det\_stat}  & \textsc{epn}                 & Detection statistic, \textsc{epn} fit    &           & No                 \\
\textsc{dbl\_det\_stat}  & \textsc{dbl}                 & Detection statistic, \textsc{dbl} fit    &           & No                 \\
\textsc{ext\_stat}       & \textsc{ext}                 & Extent statistic, \textsc{ext} fit       &           & Yes                \\
                         &                              & Significance of \textsc{ext} fit over \textsc{pnt} fit  &     &                    \\  
\textsc{epn\_stat\_ext}  & \textsc{epn}                 & Extent statistic, \textsc{epn} fit       &          & No                 \\
                         &                              & Significance of \textsc{epn} fit over \textsc{ext} fit  &     &                    \\
\textsc{epn\_stat\_pnt}  & \textsc{epn}                 & Extent statistic, \textsc{epn} fit       &          & No                 \\
                         &                              & Significance of \textsc{epn} fit over \textsc{pnt} fit  &     &                    \\
\textsc{dbl\_stat}       & \textsc{dbl}                 & Double statistic, \textsc{dbl} fit       &          & No                 \\
                         &                              & Significance of \textsc{dbl} fit over \textsc{pnt} fit  &     &                    \\
\hline
\end{tabular}
\end{table*}

Again, we expect a point source to have a $0$ value or low values (to be defined)
 of \textsc{epn\_stat\_det}, \textsc{epn\_stat\_pnt}, and \textsc{dbl\_stat}.

\section{Testing \xaminF~with simulations}
\label{sec:sim}

We use a dedicated suite of simulations to test \xaminF~performance; 
all simulations are carried out in $[0.5-2]~\kev$; details of the simulation
(count rate, core radii of input clusters, and so on) are given
in Tables \ref{tab:simc}, \ref{tab:simcc}, and \ref{tab:simd}.

\subsection{Simulated sources}
\label{sec:ovimages}

Creating images of simulated clusters consists of the following steps.
We start by creating simulated \emph{ideal} $5\deg~\times 5\deg$ photon
images, without any background, resolved AGNs, and instrumental effect, of clusters:
the images represent a perfect X-ray sky observed with an infinite exposure time, where
only the sources of interest are present.
In practice this is achieved by using a very large
exposure time, $10^6$ seconds, to have enough photons in the image, and
re-scaling the image by this exposure time so that each pixel represents
the number of photons per second in a $2.5\arcsec\times 2.5\arcsec$ region of sky
seen by a $1~\mathrm{cm}^2$ perfect detector.
The number of actual photons for each cluster in the image is drawn
from a Poisson distribution with mean given by the cluster count
rate $\times 10^6$ seconds.
We assume input clusters to be described by a $\beta$ model \citep{cavaliere76}
with $\beta=2/3$; so they are specified by their core radius in $\arcsec$
and their count rates in $\mathrm{count/s}$.

Input clusters are placed on a $5\deg\times 5\deg$ image; we build one image per
core radius$-$count rate combination.
These images are tiled with the same tiling scheme of the XXL survey
(XXL Paper I), with pointings displaced by $20\arcmin$ from each other;
clusters are placed either at the geometrical centre  of a pointing or at
$10\arcmin$ from it, so that a large fraction of them is covered by more
than one pointing (see Figure \ref{fig:mosaic_mrfilter}).
It is important to note that the geometrical centre of a pointing is not the same
as its centre of  optics, i.e. the point of maximum telescope sensitivity; these two
points may be displaced from one another by as much as $1\arcmin$, so the space
distribution of the input sources samples the $0\arcmin-1\arcmin$ and
$9\arcmin-11\arcmin$ off-axis range.

Count rates of $0.005, 0.01, 0.05, 0.1~\mathrm{count/s}$ are used;
Core radii of $10, 20, 50\arcsec$ are used; for $0.05~\mathrm{count/s}$ we also consider
$40\arcsec$ for the reasons explained in Subsection \ref{subsec:poor}.

Double sources and clusters contaminated by a central AGN
are simulated in a similar way.
For contaminated clusters the AGN is just put in the central pixel of the cluster;
although this is a very simplified model, as in real clusters the AGN may not be
at the centre, it is sufficient for our purpose, and  testing \xaminF~with more realistic
simulations will be left to another paper.  
For double sources the AGNs are put at two pixels separated by a prescribed distance.
For contaminated clusters the AGN has a count rate that is double that of  the
cluster; for double sources the AGNs have the same count rates and are
placed at $6\arcsec$ and $12\arcsec$ (we recall that the XMM-Newton PSF
on-axis is $\approx 6\arcsec$).
The AGNs at the centre of the clusters and those in pairs close
on the sky are in addition to those which constitute the
resolved AGN background described in Subsection \ref{subsec:bkgmodel}.

The choice of count rates and core radii for simulated clusters is driven
by the need to conveniently bracket the corresponding values of
observed XXL clusters: a count rate range $[0.005-0.1]~\mathrm{count/s}$
translates to $[50-1000]$ counts on-axis for a nominal XXL
$10000\mathrm{s}$ exposure time and is appropriate
since typical XXL clusters have at most a few hundred  photons (XXL Paper I);
a core radius range $[10-50]\arcsec$ is appropriate as the typical core
radius of XXL clusters is $\approx 20\arcsec$ (XXL Paper II).

The features of these simulated sources are summarised in tables
\ref{tab:simc}, \ref{tab:simcc}, and \ref{tab:simd}.

\subsection{Realistic background model}
\label{subsec:bkgmodel}

The background needed by the simulated images consists of four components.

The first componenent is the 
particle background (unvignetted): it comes from the spectrum accumulated in
$200~\mathrm{ks}$ exposures with the
EPIC filter wheel in closed position so as to not have X-ray photons pass through
and selected not to have flares.
The background is then created with a uniform spatial distribution over each CCD and the
energy distribution follows that spectrum;

The second component is the
resolved AGN background (vignetted): it is taken from the \lognlogs~relation of
\citet{moretti03}; AGNs are uniformly distributed in a $25 \desqg$ FoV and hence
they may fall near clusters or close to each other.
The flux limit is $10^{-16}~\funits$ which gives $80536$ AGNs; photons are
distributed in $[0.5-2]~\kev$ according to a power law with index $1.9$ and
with Poisson noise.
The flux is corrected for Galactic absorption according to \citet{morrison83};

The third component is the
diffuse photon background (vignetted): the model adopted is taken from
\citet{snowden08}.
Galactic photon background is described by
two MEKAL models \citep{mewe85,mewe86,liedahl95}
for the Local Hot Bubble (LHB) and the cold halo respectively,
both with a plasma temperature of $0.1~\kev$, and one MEKAL model
for the hot halo with a plasma temperature of $0.25~\kev$.
Unresolved extra-galactic sources are modelled by a power law index $1.46$.
All diffuse photon background components except the LHB are corrected for
Galactic absorption assuming a column density $1.2\times 10^{20}~\mathrm{cm}^2$.
Calculations are performed with
XSPEC, described in \citet{arnaud96};

The fourth component is the
residual contamination from soft protons
\citep[SP:][]{read03,deluca04,leccardi08,snowden08} (vignetted): it is due to
interactions between particles accelerated in the Earth's magnetosphere that reach
the detector and simulate the effect of a photon.
Although the most severe episodes can be easily identified and removed, some
residual contamination may remain in supposedly `clean' observations and must
be taken into account.
To model residual SP contamination we again follow  \citet{snowden08} who model
it as a single power law with index $\approx 0.9$, although other parameterizations are
possible (\citealt{leccardi08} adopt a double power law broken at $5~\kev$).

The relative normalisations of the different diffuse photon background
components are taken from Table 2 of \citet{snowden08};
we checked, by generating $100$ simulated pointings containing only
backgrounds (both photon and particle) that the mean background per pixel
is $\approx 10^{-5} \mathrm{photons/sec/pixel}$, compatible with the measured
XXL background level (Figure 7 of XXL Paper I).
We note that our background modelling is not entirely realistic as it
is neither time nor position dependent (as real X-ray background is);
however, we think that it is sufficient for our purposes of validating
\xaminF~using simplified analytical simulations.

\subsection{Making realistic simulated XMM-Newton images}
\label{sec:images}

We simulate images as follows:

\begin{enumerate}

\item Each XMM pointing is simulated independently;

\item For both the simulated clusters and three of the background components
described in Subsection \ref{subsec:bkgmodel} (the particle background is
included at a successive time) a perfect event list is created and fed to a
dedicated IDL routine, which creates a realistic XMM-Newton image introducing the
relevant instrumental effects;

\item In all cases an exposure time of $10~\mathrm{ks}$ and THIN filters are
assumed so as to reproduce the real XXL observing conditions;

\item For the photons in the perfect event lists which come from the simulated
clusters, an energy of $1~\kev$ (appropriate for the low-mass clusters
preferentially observed by XXL) is assumed;

\item
Other photons from resolved AGNs, diffuse photon background, and soft protons
are distributed in energy according to the models described in
Subsection \ref{subsec:bkgmodel};

\item Galactic absorption is corrected for according to \citep{morrison83};

\item The routine assumes as input an ideal event list coming from a perfect
detector with an effective area $1387.71~\mathrm{cm}^2$, appropriate for
pn THIN filters at $\approx 1.5~\kev$, where the XMM-Newton sensitivity is maximum;

\item `Blurred' event lists are created from the perfect input event
lists, one for each EPIC detector; they include instrumental effects such
as vignetting and blurring in energy and position;

\item To create these blurred event lists, input photons are reshuffled in
position and energy and/or thrown away according to their initial position and
energy.
This reshuffling takes into account the different effective areas and PSF
distortions, both strongly energy and position dependent, of the three EPIC
detectors;

\item
Particle background is added according to the model described in
\ref{subsec:bkgmodel};

\item
Images in $[0.5-2]~\kev$ at $2.5\arcsec/\mathrm{pixel}$ (one for each EPIC detector),
to oversample the XMM-Newton PSF, are then created along with the corresponding
exposure maps; the format is the same as the real XMM-Newton images;

\item Images are created so as to reproduce the real XXL tiling with pointings
are spaced by $20\arcmin$ (XXL Paper I).
To cover $25~\deg^2$ of sky with this tiling $256$ overlapping pointings are needed.
Exposure maps and detector masks are created as well;

\item
After creating an event list for each pointing, event lists from neighbouring
pointings are re-projected to a common point on the sky to make a \atile~tile per
detector; exposure maps are also re-projected and \atile~tiled exposure maps are
created.

\end{enumerate}

\begin{figure*}
        \centering
        \includegraphics[width=\linewidth]{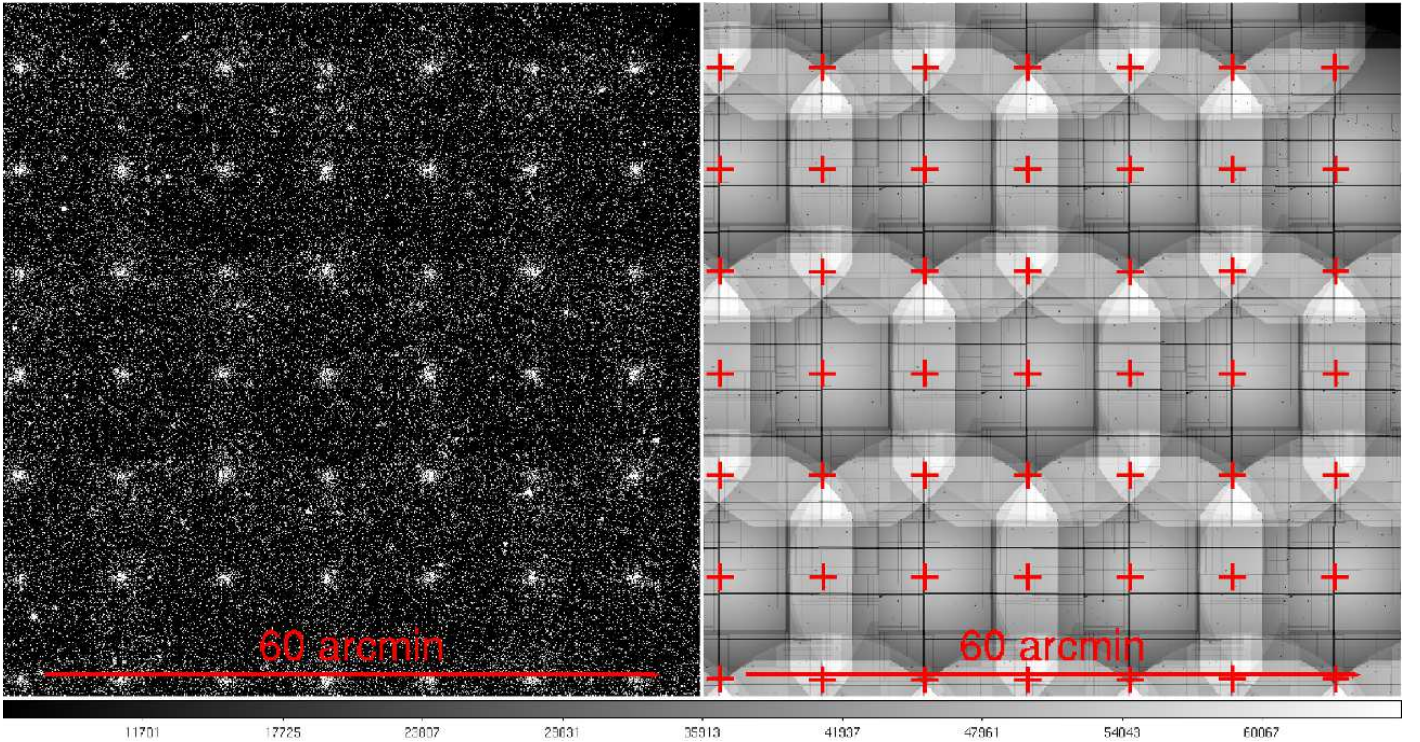}
        \caption{
        Examples of a simulated tile and its corresponding tiled exposure map.
        Left panel: Mosaic (MOS1$+$MOS2$+$pn) raw X-ray tile, comprising $\approx 25$
        overlapping pointings.
        Sources in the tile include both clusters (larger sources regularly spaced by $10\arcmin$)
        all with count rates of $0.1~\mathrm{count/s}$ and core radius $20\arcsec$,
        and resolved AGNs inserted at random positions; background is added.
        A wavelet smoothed image of the same area is shown in Figure \ref{fig:mosaic_mrfilter}.
        Right panel: Combined (MOS1$+$MOS2$+3.1\times$pn) tiled exposure map of the same region.
        Red crosses show the cluster positions.
        A nominal XMM-LSS exposure time $10~\mathrm{ks}$ for each instrument is assumed in both panels.
        The exposure map is in MOS units: the pn exposure map is corrected for the larger pn effective area by
        multiplying it by $3.1$ (see text) so the combined exposure map has an exposure time $\approx 50~\mathrm{ks}$
        at the pointing centres.
        The area is the same as that shown in Figure \ref{fig:particlebgd} and the exposure maps in the right panels
        of both figures are the same.
        }

        \label{fig:simmosaics}
\end{figure*}

\begin{figure}
        \centering
        \includegraphics[width=\linewidth]{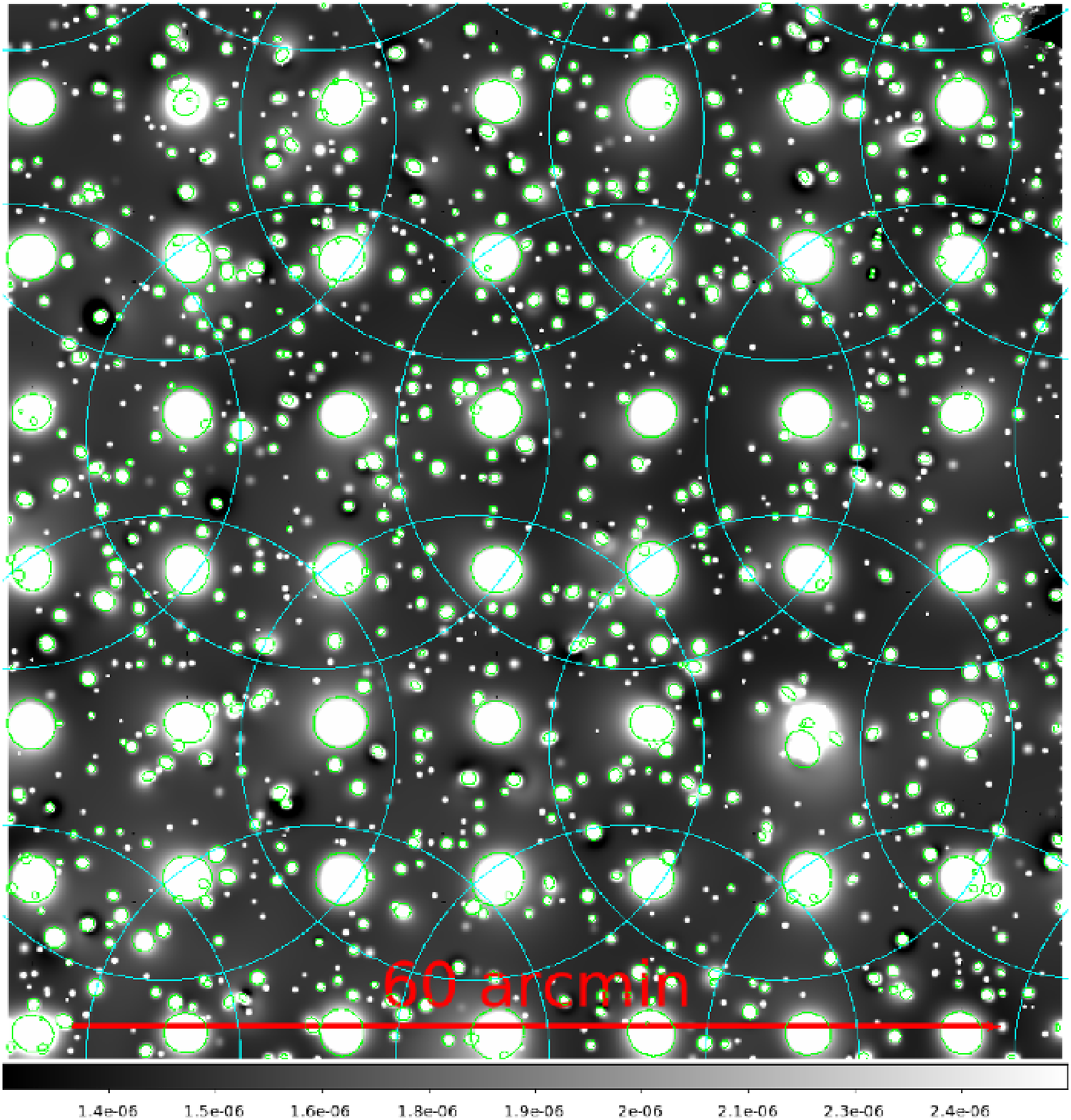}
        \caption{
        \atile~combined (MOS1$+$MOS2$+$pn) tile smoothed with \texttt{MR/1}
        on which preliminary source detection with \sex~has been run.
        The figure is the smoothed version of the left panel of Figure
        \ref{fig:simmosaics}. 
        Green circles indicate sources tentatively identified by \sex.
        Detected sources include clusters (larger sources regularly spaced by $10\arcmin$)
        with count rates of  $0.1~\mathrm{count/s}$ and core radius $20\arcsec$, and resolved AGNs
        inserted at random positions (smaller sources irregularly spaced).
        The exposure time is $10~\mathrm{ks}$ and background is added.
        Cyan circles represent the XMM pointings and are $15\arcmin$ in radius; the pointing geometrical
        centres at which the circles are drawn are not the same as the centres of the optics
        (points of maximum telescope sensitivity).
        }
        \label{fig:mosaic_mrfilter}
\end{figure}

Figure \ref{fig:simmosaics} shows an example of a simulated \atile~MOS1$+$MOS2$+$pn
tile of several bright simulated clusters (all with count rate $0.1~\mathrm{count/s}$
and core radius $20\arcsec$) and its corresponding tiled exposure map.
The tile includes $\approx 22$ overlapping pointings; the red crosses in the right panel
show the cluster positions; the wavelet smoothed image of the same area is shown in
Figure \ref{fig:mosaic_mrfilter}.

Figure \ref{fig:mosaic_mrfilter} shows a \atile~tile smoothed with the \texttt{MR/1}
wavelet smoothing code on which preliminary source detection with \sex~has been run; the
figure is the smoothed version of the X-ray raw tile shown in the left panel of
Figure \ref{fig:simmosaics}.

\subsection{PSF model}
\label{sec:psf}

\begin{figure}
        \centering
        \includegraphics[width=\linewidth]{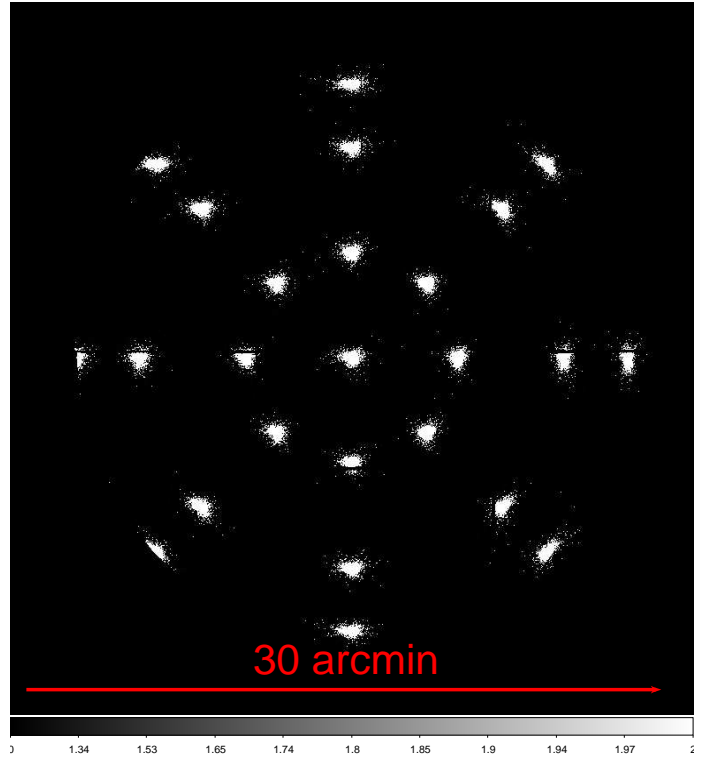}
        \caption{PSF distortion of the MOS2 detector for a point source with a
        $1~\kev$ energy put at $0\arcmin$, $5\arcmin$, $10\arcmin$, and $13\arcmin$
        from the centre of the image.
        No blurring in energy or vignetting is introduced here to show more clearly the
        effects of PSF distortion.}

        \label{fig:ELLBETA_m2}
\end{figure}

Our PSF model is the latest ELLBETA model for XMM-Newton described in \citet{read11}.
More specifically we use the SAS task \texttt{psfgen} to generate
PSF images at $1.1\arcsec/\mathrm{pixel}$ and at a range of energies, off-axis
angles, and azimuthal angles for each EPIC detector.
These images are stored together in a FITS file and our program computes a random shift from its nominal position 
for each
event, first by interpolating between
the images according to the event energy and position and then sampling a random
shift from the interpolated image;  PSF images for the appropriate detector
are used each time. 
An example is shown for the MOS2 detector in Figure \ref{fig:ELLBETA_m2};
the figure shows how a point source with a $1~\kev$ energy put at $0\arcmin$,
$5\arcmin$, $10\arcmin$, and $13\arcmin$ from the centre of the image and
at several azimuthal angles is distorted by the XMM-Newton optics.
No blurring in energy or vignetting was introduced in making the figure to show more
clearly the effects of PSF distortion (but of course they {are} introduced in
the simulated images).

Similar figures could be drawn for MOS1 and pn and they would show equally strong
off-axis and azimuthal dependent distortions, though the shapes of the distortions
would be different in each detector.  

\section{Results}
\label{sec:ov}

\subsection{Recovering clusters}
\label{sec:clusters}

In the following we  compare the relative performance of using single pointings
and  using all available overlapping pointings.
We will loosely speak of `tiles versus pointings' but it should be remembered that
\atile~tiles are used only for source detection by \sex~and that multiple
overlapping pointings are used for the actual fit.
In addition, even in the case of single pointings, fits have been performed with the new
\xaminF~incorporating all other improvements (new PSF model, new fits, etc.), so
the results are not exactly the same as those that would be obtained by \xaminP.

\subsubsection{Tiles vs. single pointings}

\begin{figure*}
        \centering
        \begin{subfigure}[b]{0.45\linewidth}
                \includegraphics[width=\linewidth]{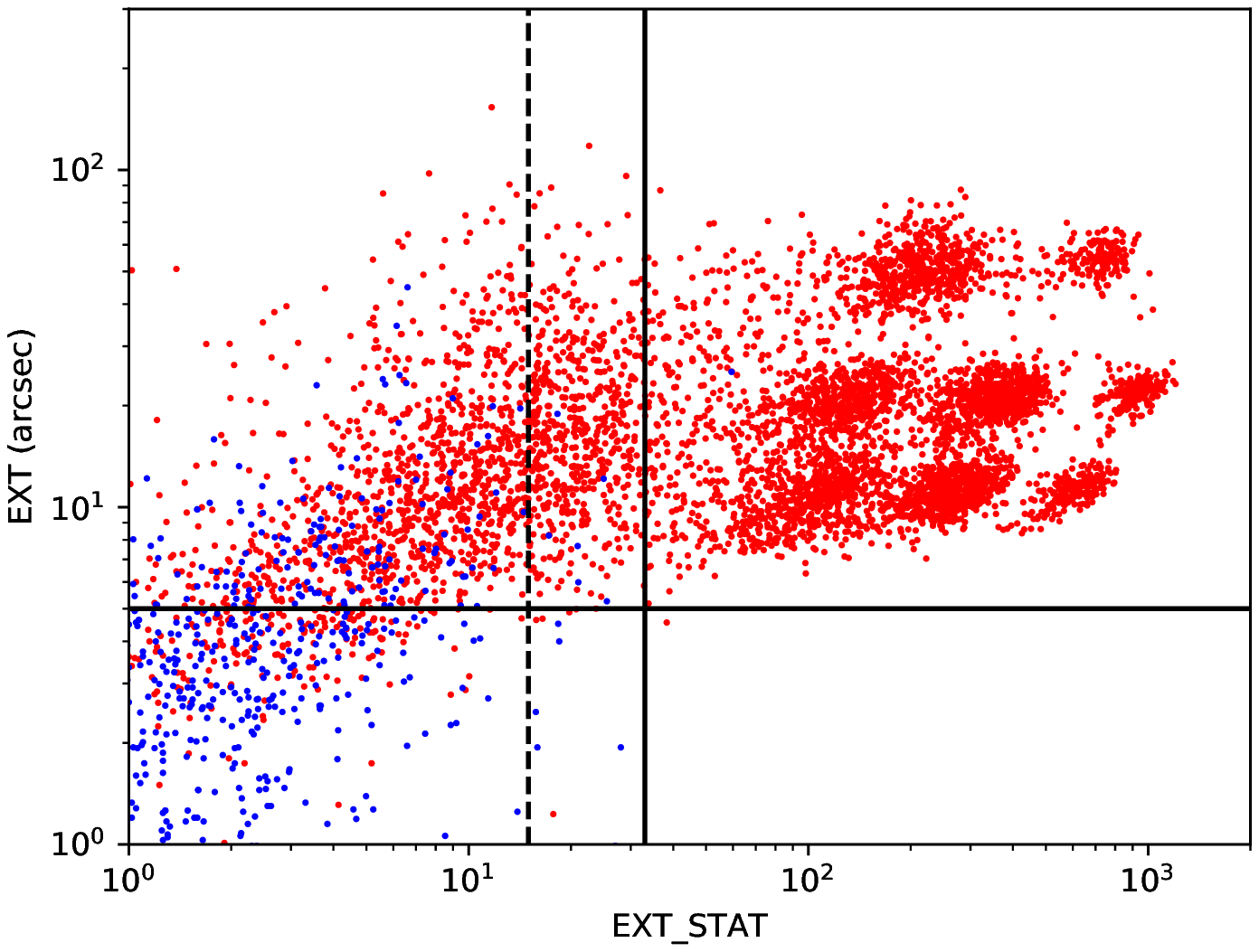}
        \end{subfigure}
        \begin{subfigure}[b]{0.45\linewidth}
                \includegraphics[width=\linewidth]{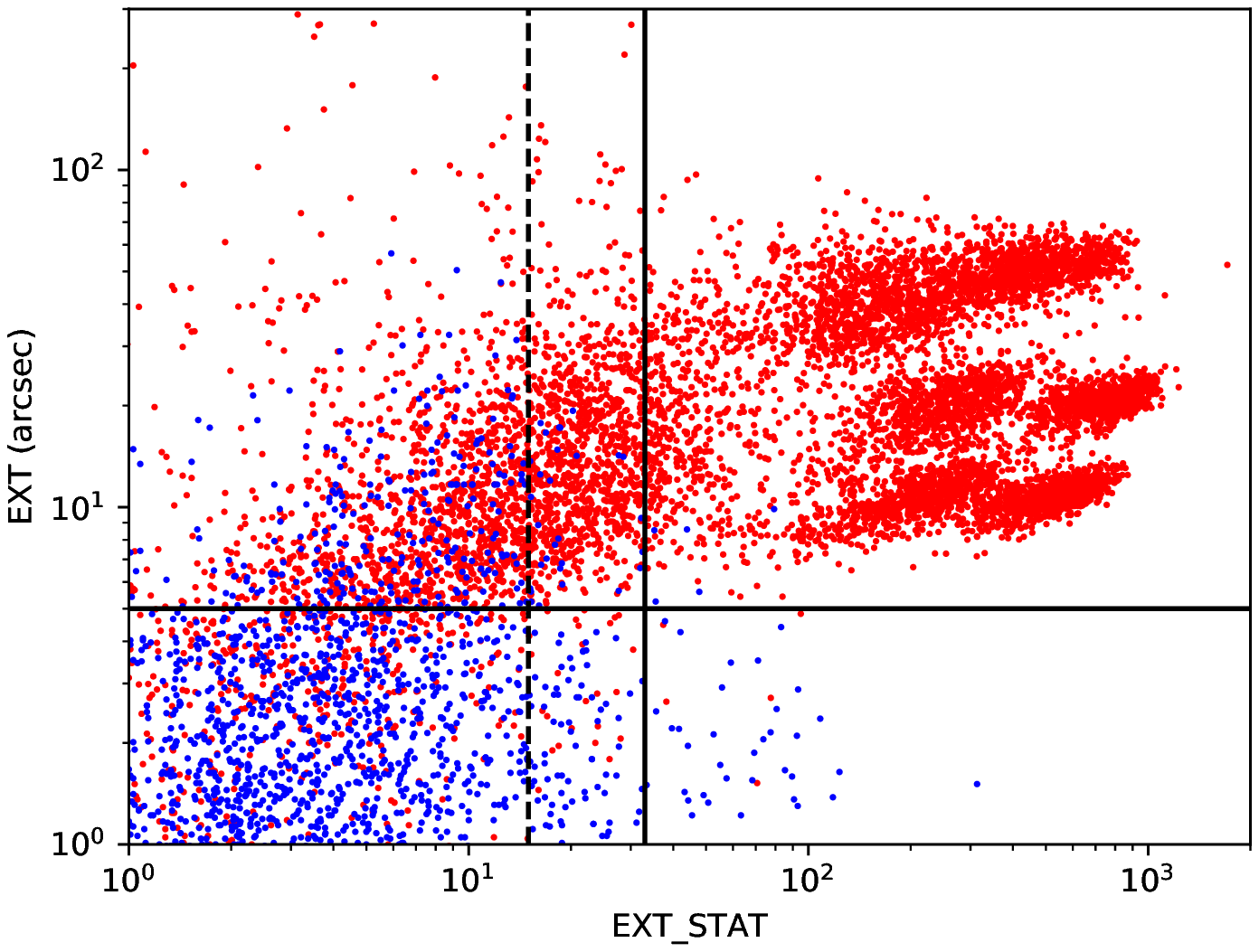}
        \end{subfigure}

        \begin{subfigure}[b]{0.45\linewidth}
                \includegraphics[width=\linewidth]{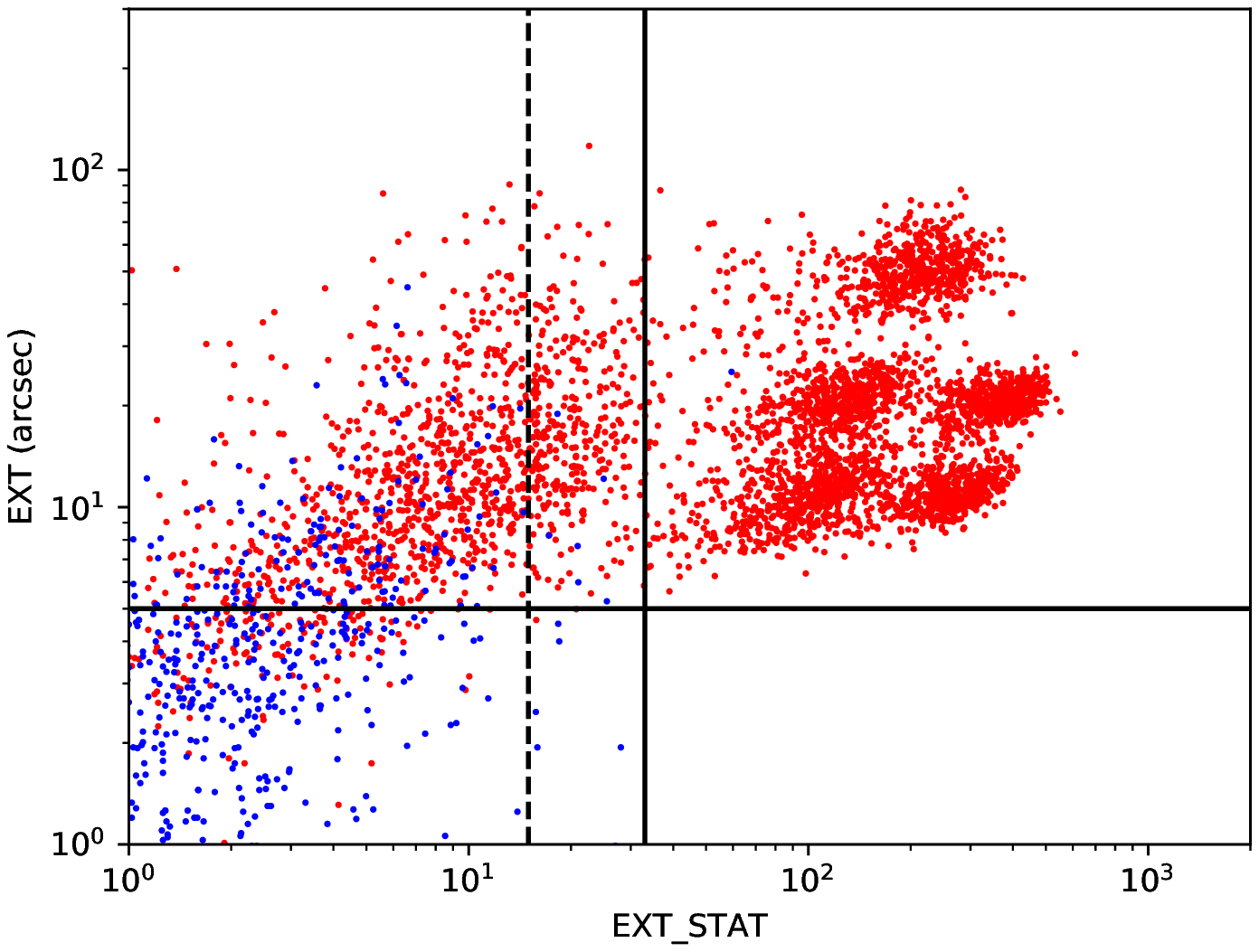}
        \end{subfigure}
        \begin{subfigure}[b]{0.45\linewidth}
                \includegraphics[width=\linewidth]{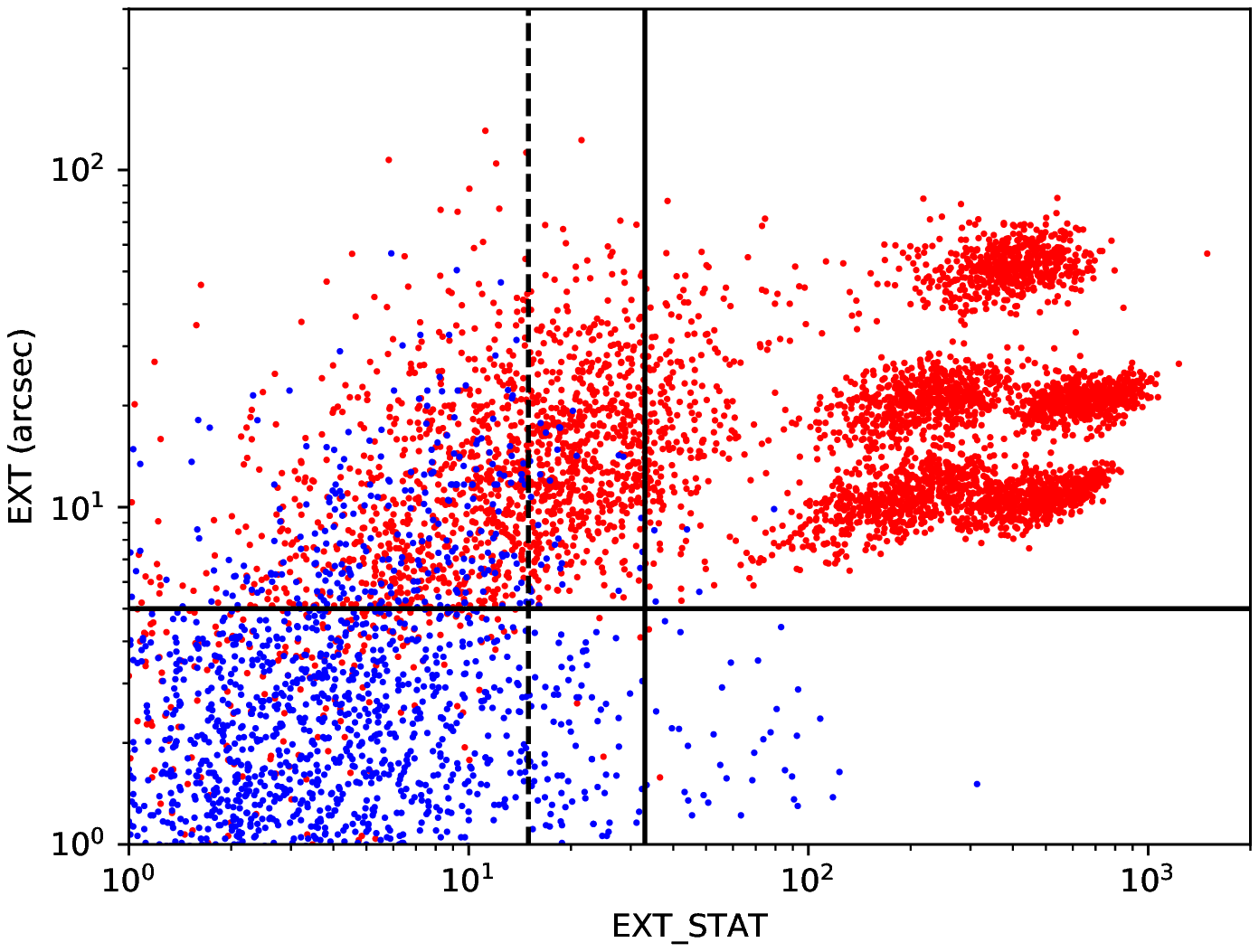}
        \end{subfigure}

        \caption{\textsc{ext} vs. \textsc{ext\_stat} plane for recovered clusters (red) and AGNs (blue).
        The continuous line in each panel at $\textsc{ext\_stat}=33$ shows the cut for the C1 selection
        and the dashed line at $\textsc{ext\_stat=15}$ shows the cut for the C2 selection.
        The continuous line at $\textsc{ext}=5\arcsec$ shows the cut in extent for both the C1 and C2 selections.
        If a source is found in more than one tile or pointing, the detection with the highest
        value of \textsc{ext\_stat} is shown.
        Panel a (top left): single pointings; all sources.
        Panel b (top right):  tiles; all sources.
        Panel c (bottom left): single pointings; off-axis angle $>9\arcmin$.
        Panel d (bottom right): tiles; off-axis angle $>9\arcmin$.
        }

        \label{fig:ext_vs_ext_like_ov}
\end{figure*}

Figure \ref{fig:ext_vs_ext_like_ov} shows the recovered core radius (\textsc{ext})
versus the extent statistic (\textsc{ext\_stat}) of the \textsc{ext} fit for
recovered clusters (red points) and AGNs (blue points).
The panels in the figure all show cluster detections within $37.5\arcsec$ of an
input cluster (except for clusters with input core radius $40\arcsec$ where the
correlation radius is $50\arcsec$, and clusters with input core radius $50\arcsec$
where the correlation radius is $60\arcsec$).
If the same input cluster is detected more than once because it falls on more than
one tile or one pointing, the detection with the highest value of \textsc{ext\_stat}
is shown.
AGN detections are taken from a $25 \deg^2$ simulation containing only
AGNs randomly distributed in the field and with a flux distributed according to
a \citet{moretti03} \lognlogs~relation; detections from these simulations within
$6\arcsec$ of an input AGNs are reported.
The figure shows all detections (top panels)
and only the detections of input clusters placed at off-axis angle $>9\arcmin$  measured
from the centre of the optics, not the geometrical centre, from all pointings
that cover them (bottom panels).
We chose $>9\arcmin$ even though clusters are either at the centre or displaced by
$>10\arcmin$ from the geometrical centre because the centre of the optics can be
offset from it by as much as $\approx 1\arcmin$.
In both cases in Figure \ref{fig:ext_vs_ext_like_ov} the left panels pertain to
the case where only single pointings  are used and the right panels pertain to
the case where tiles are used.
The lines show the C1$/$C2 selection criteria defined in P06.
The figure shows that the separation between clusters (red points) on the
one hand and AGNs (blue points) on the other is very clear; the main point is
that the C1$/$C2 selection criteria defined by P06 are still valid and do
not need to be revised in view of the changes intervened in both the PSF model
and the wavelet smoothing program between \xaminP~and \xaminF, currently not
tested beyond $10~\mathrm{ks}$.

\begin{figure}
        \centering

        \resizebox{\hsize}{!}{\includegraphics{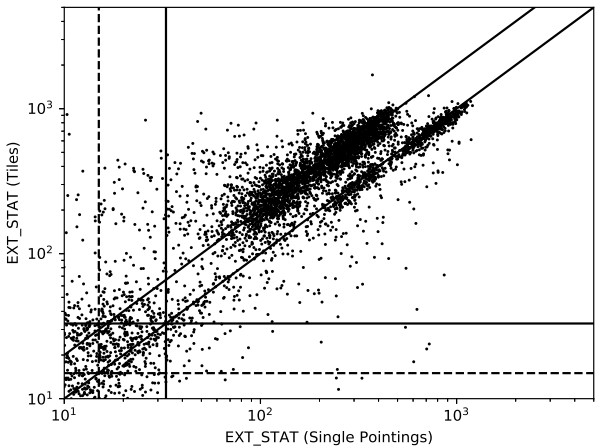}}
        \caption{\textsc{ext\_stat} using tiles vs. \textsc{ext\_stat} using single pointings.
        The lines at $\textsc{ext\_stat}=33$ show the C1 selection criterion, the dashed lines at
        $\textsc{ext\_stat}=15$ the C2 selection criterion.
        Also shown are the line of equality, where on-axis sources fall, and the line
        \textsc{ext\_stat}$_\mathrm{Tiles} = 2\times$ \textsc{ext\_stat}$_\mathrm{Pointings}$, where most
        sources off-axis fall, showing the \textsc{ext\_stat}$_\mathrm{Tiles}\approx~2$ \textsc{ext\_stat}$_\mathrm{Pointings}$.
        }

        \label{fig:ext_like_tiles_vs_pointing}

\end{figure}

Other trends are apparent in Figure \ref{fig:ext_vs_ext_like_ov}.
Comparing the two left panels, i.e. the detections using single pointings (top,
all detections; bottom, only detections of sources at off-axis angle $>9\arcmin$
from all pointings) we see that the three red clouds at the top, corresponding to
the highest values of \textsc{ext\_stat}, disappear at the bottom; these are the
detections of the sources on-axis, which are detected with the highest significance.
Comparing the two right panels, i.e. detections using tiles (top, all detections;
bottom, only detections of sources at off-axis angle $>9\arcmin$ from all
pointings); however,  we see that the shapes of the red clouds are similar. This
means that sources on-axis and at large off-axis are recovered at the same level of
significance, which is not the case with single pointings.
There is also an overall trend toward higher values of \textsc{ext\_stat} moving from
single pointings to tiles (compare left and right panels in Figure
\ref{fig:ext_vs_ext_like_ov}).
Comparing the two top panels we see that the isolated clouds on the left disappear on
the right, and that the distribution of \textsc{ext\_stat} becomes more uniform: sources
detected at lower \textsc{ext\_stat} in single pointings migrate toward higher values
of \textsc{ext\_stat}.
This is not surprising: we  expect that using, on average, two pointings
instead of one would lead to a doubling, on average, of \textsc{ext\_stat} for
sources at large off-axis angles.
The effect is demonstrated in Figure \ref{fig:ext_like_tiles_vs_pointing} which
shows \textsc{ext\_stat} using tiles versus \textsc{ext\_stat} using single pointings.
The figure shows the line of equality
(\textsc{ext\_stat}$_\mathrm{Tiles} = $ \textsc{ext\_stat}$_\mathrm{Pointings}$)
where sources on-axis fall, and the line
\textsc{ext\_stat}$_\mathrm{Tiles} = 2~\times$ \textsc{ext\_stat}$_\mathrm{Pointings}$
where the bulk of the off-axis sources fall, which means  that  the level of
significance roughly doubles using tiles.

\begin{figure*}
        \centering
        \begin{subfigure}[b]{0.45\linewidth}
                \includegraphics[width=\linewidth]{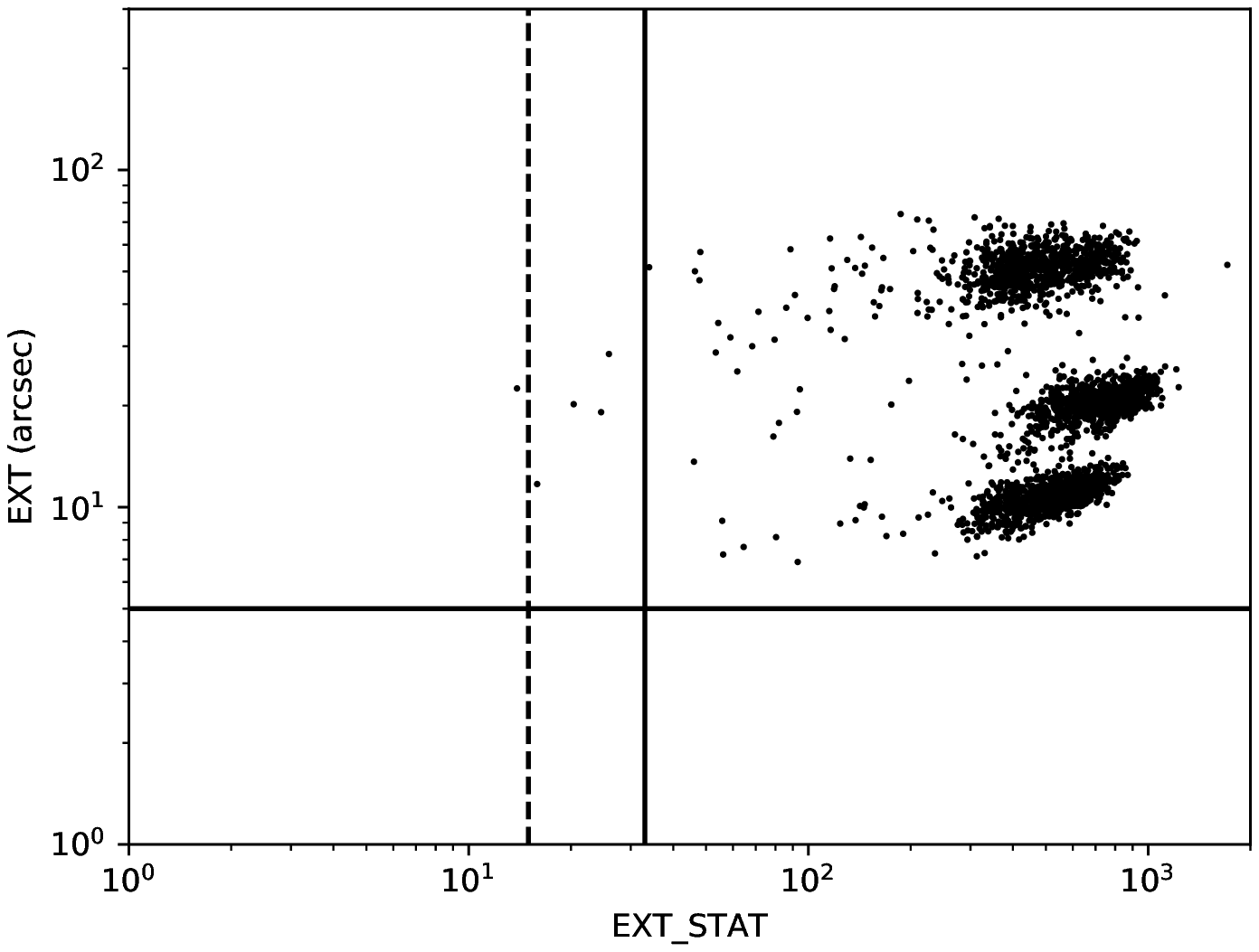}
        \end{subfigure}
        \begin{subfigure}[b]{0.45\linewidth}
                \includegraphics[width=\linewidth]{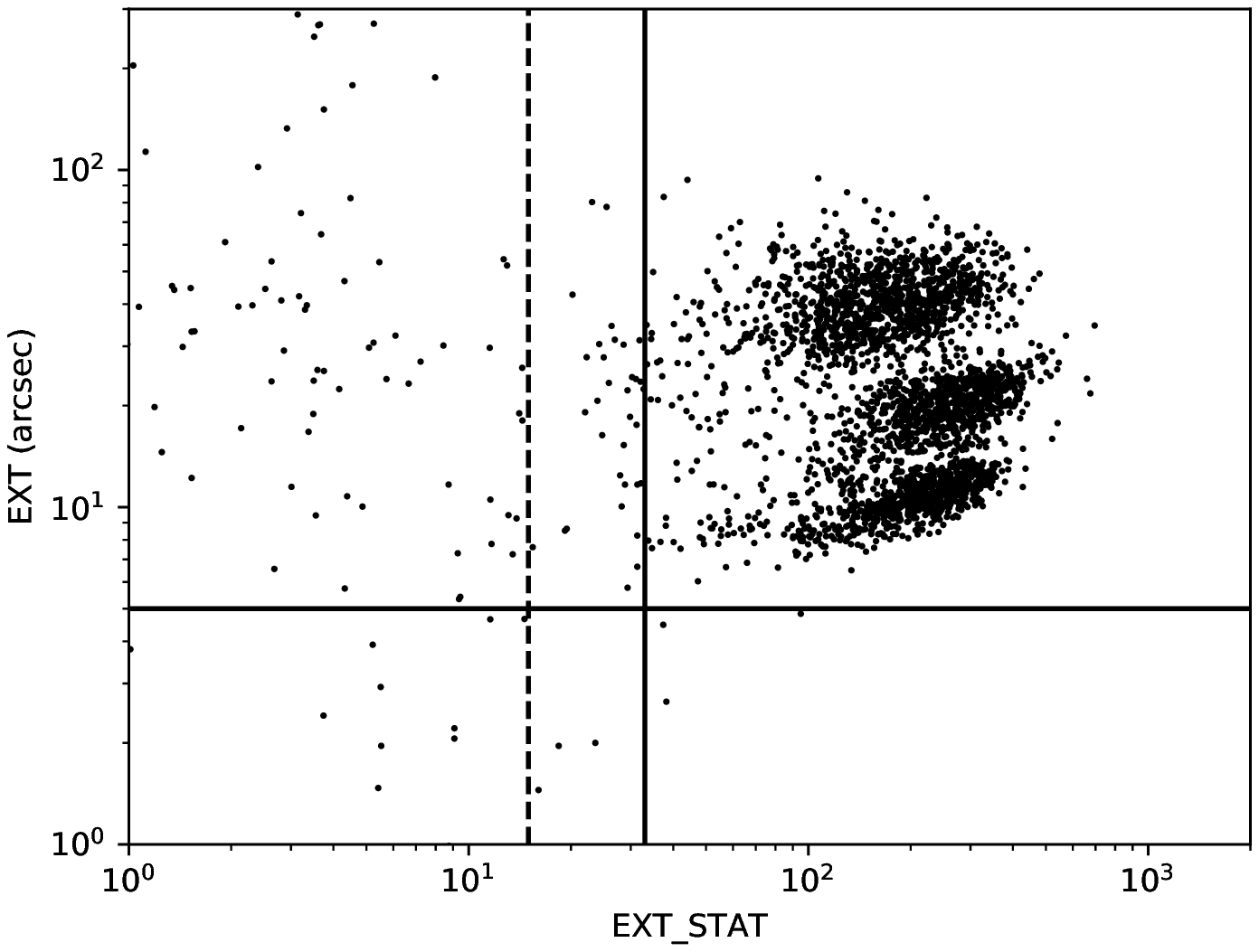}
        \end{subfigure}

        \begin{subfigure}[b]{0.45\linewidth}
                \includegraphics[width=\linewidth]{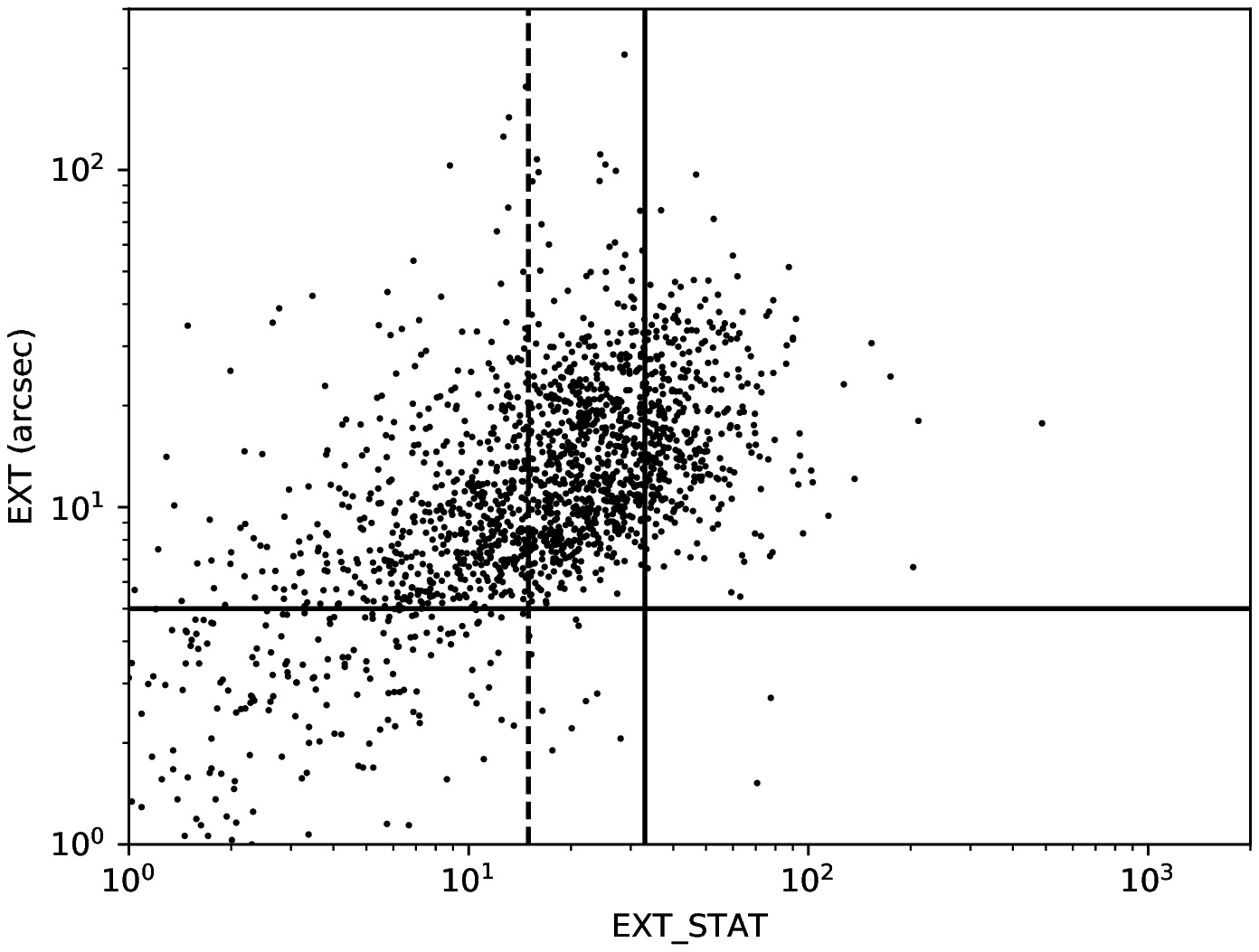}
        \end{subfigure}
        \begin{subfigure}[b]{0.45\linewidth}
                \includegraphics[width=\linewidth]{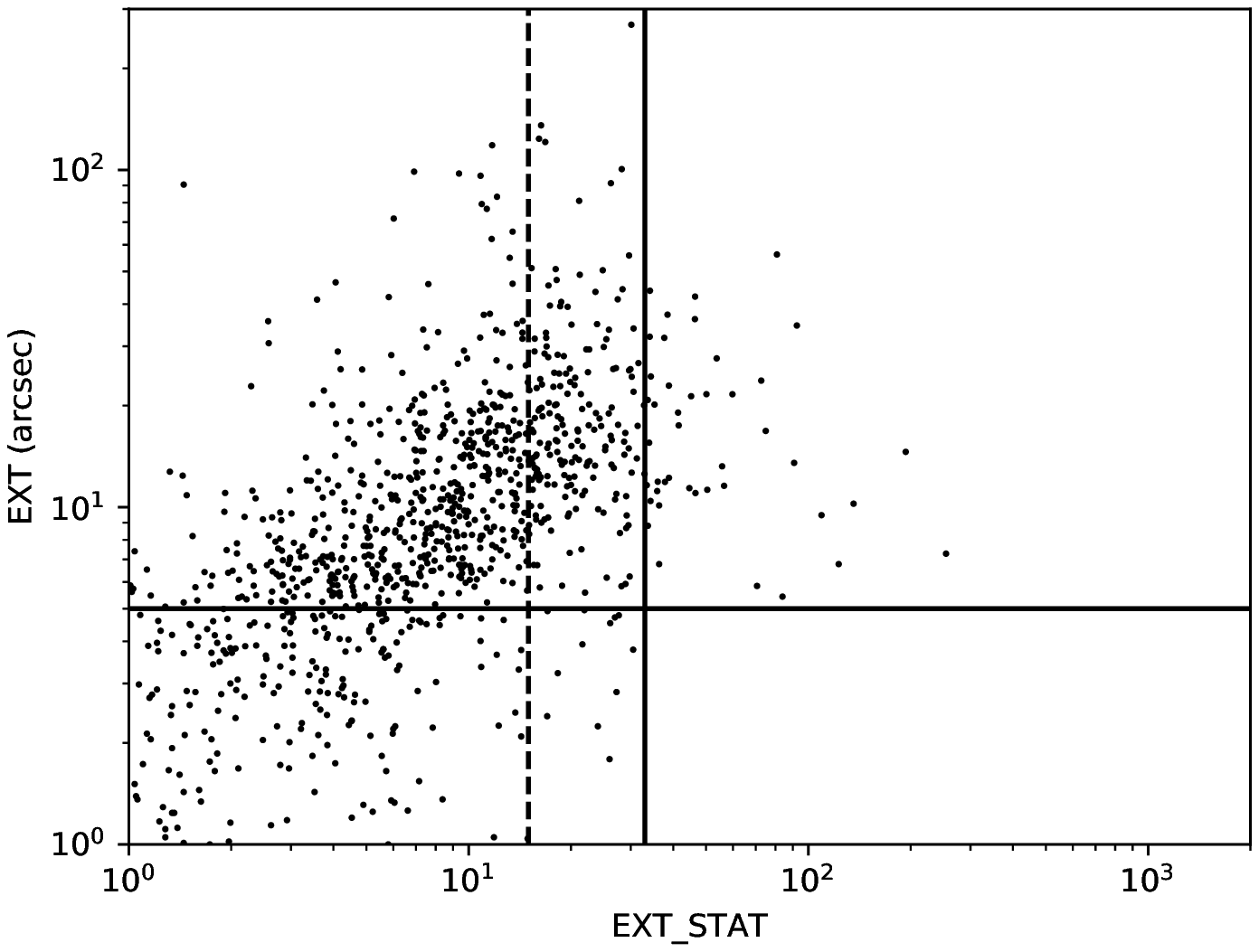}
        \end{subfigure}

        \caption{\textsc{ext} vs. \textsc{ext\_stat} plane for recovered clusters where clusters have been
        separated by input count rate.
        The continuous line at $\textsc{ext\_stat}=33$ shows the cut for the C1 selection
        and the dashed line at $\textsc{ext\_stat}=15$ shows the cut for the C2 selection.
        The continuous line at $\textsc{ext}=5\arcsec$ shows the cut in extent for both the C1 and C2 selections.
        If a source is found in more than one tile, the detection with the highest
        value of \textsc{ext\_stat} is shown.
        Panel a (top left): tiles; input count rate $0.1~\mathrm{count/s}$.
        Panel b (top right): tiles; input count rate $0.05~\mathrm{count/s}$.
        Panel c (bottom left): tiles; input count rate $0.01~\mathrm{count/s}$.
        Panel d (bottom right): tiles; input count rate $0.005~\mathrm{count/s}$.
        }
        \label{fig:ext_vs_ext_like_cr}
\end{figure*}

Figure \ref{fig:ext_vs_ext_like_ov} shows several clouds of red points both in the
case of tiles and of single pointings; their presence is just due to fact that count
rates and core radii of input clusters take a few values and are not continuously
distributed.
This is easy to see for the clouds at $\mathrm{ext}=[10, 20, 50]\arcsec$ in all
panels of Figure \ref{fig:ext_vs_ext_like_ov}, and is further seen in Figure
\ref{fig:ext_vs_ext_like_cr} which shows again the \textsc{ext} versus
\textsc{ext\_stat} plane, using tiles, of recovered clusters where clusters
have been separated by input count rate.

\subsubsection{Detection probabilities}

We now show the detection probabilities of our simulated clusters.
Detection probabilities are computed, for each count rate$-$core radius
combination, by dividing the number of detected C1 or C2 clusters by the number of
input clusters.
Detected clusters are found by matching a detection (C1 or C2) to the nearest
input cluster using correlation radii $[37.5,37.5,50.0,60.0]\arcsec$ for
input core radii $[10,20,40,50]\arcsec$; unmatched detections are not considered.
If an input cluster is found in more than one tile or pointing, only the detection with the
highest value of $\textsc{ext\_stat}$ is considered; therefore, if an input cluster
is found both as a C1 and a C2, only the C1 detection is considered.
In computing the probabilities we have removed input clusters and detections which
were too close to the border of the simulated $25\desqg$ sky patch to avoid border
effects; this removal reduces the number of input cluster from 841 to 784.
Tables \ref{tab:c1selfun_all}, \ref{tab:c1selfun_oa_gt9}, \ref{tab:c2selfun_all},
and \ref{tab:c2selfun_oa_gt9} list the C1 and C2 detection probabilities for
simulated clusters as a function of input count rate and core radius both using
tiles and using single pointings.
Tables \ref{tab:c1selfun_all} and \ref{tab:c2selfun_all} list the probabilities
across the whole XMM FoV and Tables \ref{tab:c1selfun_oa_gt9} and
\ref{tab:c2selfun_oa_gt9} list the probabilities for input clusters which fall at
an off-axis angle $>9\arcmin$ on all pointings that cover them to better asses the
benefits of using tiles for sources at large off-axis angles.

\begin{table*}
\caption{C1 detections for all sources; spurious cluster detections are not reported.
Total input clusters: 784.}

\label{tab:c1selfun_all}
\centering
\begin{tabular}{lcrlrl}

\hline
Input count rate   & Input core radius & C1 Detections & Fraction    & C1 Detections          & C1 Fraction \\
$\mathrm{count/s}$ & $\arcsec$         & tiles         & tiles       & single pointings       & single pointings \\
\hline

0.005 & 10 & 14 & 0.018 & 7 & 0.009 \\
0.005 & 20 & 14 & 0.018 & 1 & 0.001 \\
0.005 & 50 &  5 & 0.006 & 0 & 0.0 \\

0.01 & 10 & 160 & 0.204 & 74 & 0.094 \\
0.01 & 20 & 139 & 0.177 & 62 & 0.079 \\
0.01 & 50 &  13 & 0.017 &  6 & 0.008 \\

0.05 & 10 & 762 & 0.971 & 704 & 0.898 \\
0.05 & 20 & 779 & 0.994 & 763 & 0.973 \\
0.05 & 40 & 766 & 0.977 & 708 & 0.903 \\
0.05 & 50 & 310 & 0.395 & 573 & 0.731 \\

0.1 & 10 & 775 & 0.988 & 737 & 0.940 \\
0.1 & 20 & 783 & 0.999 & 783 & 0.999 \\
0.1 & 50 & 780 & 0.995 & 763 & 0.973 \\

\hline
\end{tabular}
\end{table*}

\begin{table*}
\caption{C1 detections for sources with input off-axis angle $>9^{\prime}$;
spurious cluster detections are not reported.
Total input clusters: 588.}

\label{tab:c1selfun_oa_gt9}
\centering
\begin{tabular}{lcrlrl}

\hline
Input count rate   & Input core radius & C1 Detections & Fraction    & C1 Detections          & C1 Fraction \\
$\mathrm{count/s}$ & $\arcsec$         & tiles         & tiles       & single pointings       & single pointings \\
\hline

0.005 & 10 &  7 & 0.012 & 1 & 0.002 \\
0.005 & 20 & 10 & 0.017 & 0 & 0.0 \\
0.005 & 50 &  3 & 0.005 & 0 & 0.0 \\

0.01 & 10 & 102 & 0.173 & 10 & 0.017 \\
0.01 & 20 &  87 & 0.148 &  2 & 0.003 \\
0.01 & 50 &   9 & 0.015 &  0 & 0.0   \\

0.05 & 10 & 576 & 0.976 & 527 & 0.896 \\
0.05 & 20 & 586 & 0.997 & 569 & 0.968 \\
0.05 & 40 & 574 & 0.976 & 518 & 0.881 \\
0.05 & 50 & 234 & 0.398 & 390 & 0.663 \\

0.1 & 10 &  586 & 0.997   &   547 & 0.930 \\
0.1 & 20 &  587 & 0.998   &   587 & 0.998 \\
0.1 & 50 &  584 & 0.993   &   572 & 0.973 \\

\hline
\end{tabular}
\end{table*}

\begin{table*}
\caption{C2 detections for all sources; spurious cluster detections are not reported.
Total input clusters: 784.}

\label{tab:c2selfun_all}
\centering
\begin{tabular}{lcrlrl}

\hline
Input count rate   & Input core radius & C2 Detections & Fraction    & C2 Detections          & C2 Fraction \\
$\mathrm{count/s}$ & $\arcsec$         & tiles         & tiles       & single pointings       & single pointings \\
\hline

0.005 & 10 & 91 & 0.116 &  56 & 0.071 \\
0.005 & 20 & 46 & 0.059 &  35 & 0.046 \\
0.005 & 50 &  9 & 0.011 &   5 & 0.006 \\

0.01 & 10 & 282 & 0.360 & 199 & 0.253 \\
0.01 & 20 & 259 & 0.330 & 138 & 0.176 \\
0.01 & 50 &  34 & 0.043 &  34 & 0.043 \\

0.05 & 10 &  1 & 0.001   &  15 & 0.019 \\
0.05 & 20 &  1 & 0.001   &  10 & 0.013 \\
0.05 & 40 & 10 & 0.013   &  40 & 0.051 \\
0.05 & 50 & 11 & 0.014   & 121 & 0.154 \\ 

0.1 & 10 &    0 & 0.0   &   5 & 0.006 \\
0.1 & 20 &    1 & 0.001 &   0 & 0.0 \\
0.1 & 50 &    2 & 0.03  &  13 & 0.017 \\

\hline
\end{tabular}
\end{table*}

\begin{table*}
\caption{C2 detections for sources with input off-axis angle $>9^{\prime}$;
spurious cluster detections are not reported.
Total input clusters: 588.}
\label{tab:c2selfun_oa_gt9}
\centering
\begin{tabular}{lcrlrl}

\hline
Input count rate   & Input core radius & C2 Detections & Fraction    & C2 Detections          & C2 Fraction \\
$\mathrm{count/s}$ & $\arcsec$         & tiles         & tiles       & single pointings       & single pointings \\
\hline

0.005 & 10 & 62 & 0.105 & 19 & 0.032 \\
0.005 & 20 & 28 & 0.047 &  4 & 0.007 \\
0.005 & 50 &  8 & 0.014 &  0 & 0.0   \\

0.01 & 10 & 210 & 0.357 & 127 & 0.216 \\
0.01 & 20 & 186 & 0.316 &  72 & 0.122 \\
0.01 & 50 &  20 & 0.03  &   7 & 0.012 \\

0.05 & 10 &  1 & 0.017  &  14 & 0.024 \\
0.05 & 20 &  0 & 0.0    &  10 & 0.017 \\
0.05 & 40 &  7 & 0.012  &  37 & 0.063 \\
0.05 & 50 &  8 & 0.014  & 115 & 0.195 \\

0.1 & 10 &    0 & 0.0   &   5 & 0.008 \\
0.1 & 20 &    1 & 0.002 &   0 & 0.0 \\
0.1 & 50 &    2 & 0.034 &  12 & 0.020 \\

\hline
\end{tabular}
\end{table*}

The most important message from these numbers is that using tiles greatly
enhances the detection probability of the weakest clusters, which constitute the
bulk of the XXL population; for example, the detection probability of a cluster
with count rate $0.01~\mathrm{count/s}$ and core radius $10\arcsec$ doubles from
$0.094$ to $0.204$.
Even more impressive is the performance increase at large off-axis angle;
in this case, for the same values of count rate and core radius, the probability
increases from $0.017$ to $0.173$.
This is important since XXL is a blind survey so a cluster is more likely
to fall at large off-axis angles
from all XXL pointings than near the centre of one pointing. 

For example, since the XMM FoV is $15\arcmin$, the probability of a cluster falling
at off-axis angle $>10\arcmin$ is $(15^2-10^2) / 15^2 = 0.56$
whereas the probability of falling within $10\arcmin$ of a pointing
is just $10^2 / 15^2 = 0.44$

Therefore, we expect that the upcoming reprocessing of the XXL data with
\xaminF~will yield a substantial increase in new, secure cluster detections.
It is also interesting to note that using tiles the sensitivity at large off-axis
angles is not much lower than the sensitivity across the whole FoV; the detection
probability of a cluster with count rate $0.01~\mathrm{count/s}$ and core radius
$10\arcsec$ is $0.173$ at off-axis $>9\arcsec$ and $0.204$ across the whole FoV.
Therefore, we conclude that we manage to do almost as well at large off-axis angles
as  on-axis.
This should be contrasted with the single pointing case where the detection
probability drops from $0.094$ to $0.017$ for the same values of count rate and
core radii.
For larger values of count rate ($>0.05~\mathrm{count/s}$) the improvement is less
impressive since these clusters already have enough photons to be securely detected
even at large off-axis angles in single pointings; however, these clusters will not
constitute the majority of the XXL population.

From Table \ref{tab:c1selfun_all} we see that we
detect $5300$ C1 clusters; in addition, we find a total of
$288$ false detections after excluding detections too close to the border
of the simulations and detections likely to be double sources (see Subsection
\ref{sec:doubles} for more details), so the fraction of false detections
in comparison with the total number of C1 clusters detected is $\approx 3.9\%$,
confirming the high purity of the C1 selection.

\subsubsection{Poor performance of \xaminF~for 
$0.05~\mathrm{count/s}-50\arcsec$ }
\label{subsec:poor}

We must justify the poorer performance using tiles with respect to using single
pointings in the case of count rate $0.05~\mathrm{count/s}$ and core radius
$50\arcsec$.
We attribute this degradation to the large value of the box chosen to estimate the
background in \sex~(see Table \ref{tab:param}): $512$ pixels instead of $64,$ as
used in P06, which in turn was necessary because we found that using $64$ pixels
led, in the case of tiles, to too many false C1 detections; using a larger box
therefore increases sample purity.
This choice may negatively affect the detection of sources with low photon density.
A cluster with count rate $0.05~\mathrm{count/s}$ and core radius $50\arcsec$
produces $500$ counts (or less if not on-axis) at the nominal XXL
$10~\mathrm{ks}$ exposure time, spread over a large area ($50\arcsec$ in radius),
so the surface brightness is low; instead,  using a large box to estimate the background
may lead \sex~to miss these sources as their contrast above the background is too
low.
However, we do not expect this  to be a serious problem as clusters with core
radius $\approx 50\arcsec$ are rare in the first place, and  in general have higher
count rates than $0.05~\mathrm{count/s}$ as we now show using a realistic
cosmological hydro-dynamical simulation.
We used the cosmoOWLS simulation \citep{mccarthy10, lebrun14}, for which
X-ray fluxes, $R_{500c}$,  and $\theta_{500c}$ for the input halos were available;
for one of its ten realisations we found, after applying a flux cut
$>5\times 10^{-16}~\funits$ (roughly the XXL flux limit for extended sources
in $[0.5-2]~\kev$), $14260$ input halos.
Count rates were measured for these simulated halos from their X-ray fluxes by
adopting a conversion factor from flux to count rate
$8.45\times 10^{+10} \mathrm{count/s} / (\funits)$, computed with
WebSpec.
The value is appropriate for a $z=0.5$, $T=2~\kev$ plasma (typical values for XXL clusters)
 described by an APEC model and for XMM-Newton
THIN filters used in XXL observations.
Core radii $\textsc{ext}$ were measured for the same halos by their
$\theta_{500c}$ by assuming a constant ratio
$\textsc{ext}/\theta_{500c} = 0.24$ \citep{pierre17}.
We found that only $189$ halos had count rate $<0.05~\mathrm{count/s}$ and core
radius $>50\arcsec$, that is $\approx 1.3\%$ 
We conclude that the poorer performance in this regime of the pipeline will not
negatively impact its overall performance too severely and the improvement in
sample purity we obtain is worth the price.
We finally note that for count rate $0.05~\mathrm{count/s}$ and core radius
$40\arcsec$ using tiles still leads to a significant improvement in C1 detections
probability: $0.977$ compared to $0.903$ across the whole field of view and $0.976$
compared to $0.881$ for the $>9^{\prime}$ region, so the degradation affects only
the very largest and fainter clusters.
Again the use of tiles allowed us to perfectly compensate for the decrease in
detector sensitivity: the detection probabilities across the whole FoV and at
$>9^{\prime}$ are almost identical.

\subsubsection{Sample purity} 
We now check the purity of the sample of recovered C1 detections.
We look, in all simulations, for sources detected as C1 that are too far from an
input clusters to be considered  legitimate detections of an input source; as
usual, our correlation radius is $50\arcsec$ for clusters with input core radius
$40\arcsec$, $60\arcsec$ for clusters with input core radius $50\arcsec$,
and $37.5\arcsec$ in all other cases; C1 detections outside these radii are
considered spurious.
We exclude spurious C1 detections too close to the border of the $25 \desqg$
simulated sky patch to avoid border effects; we also exclude spurious C1 detections
which are probably pairs of AGNs close on the sky as these can be effectively flagged
as such (see explanation in Subsection \ref{sec:doubles}).
We find that, for all count rate$-$core radius combinations the contamination rate
defined as the fraction of spurious C1 detections divided by the number of input
clusters is at most $\approx 3\%$ confirming the P06 finding that the C1 selection
is indeed very pure and remains so when using tiles.
This depends, of course, on the count rate$-$core radius sampling used in the
simulation, which may not be the real one; we also avoided projection effects.

\subsubsection{Example}

\begin{figure*}
        \centering

        \begin{subfigure}[b]{0.33\linewidth}
                \includegraphics[width=\linewidth]{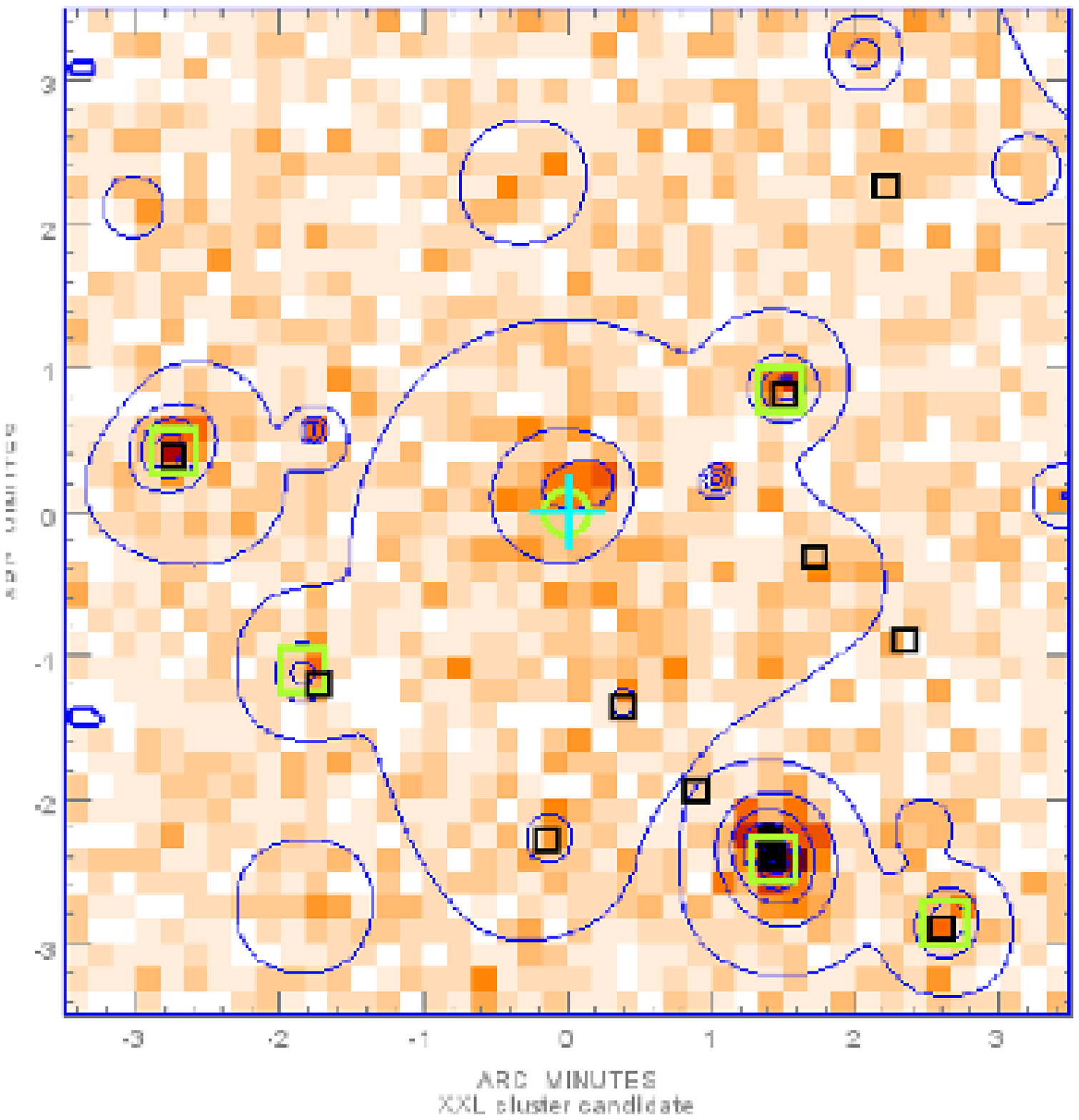}
        \end{subfigure}
        \begin{subfigure}[b]{0.33\linewidth}
                \includegraphics[width=\linewidth]{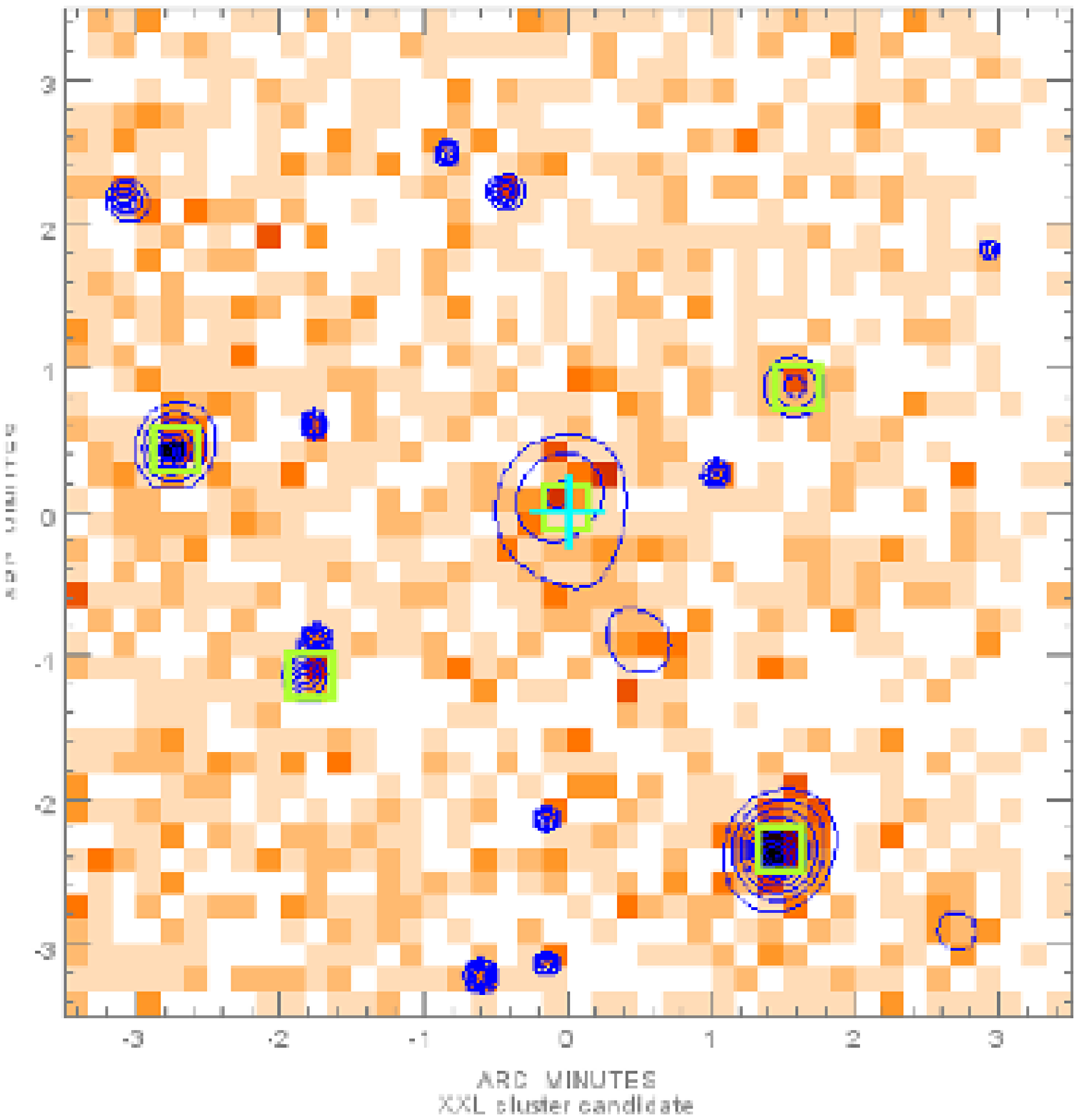}
        \end{subfigure}
        \begin{subfigure}[b]{0.33\linewidth}
                \includegraphics[width=\linewidth]{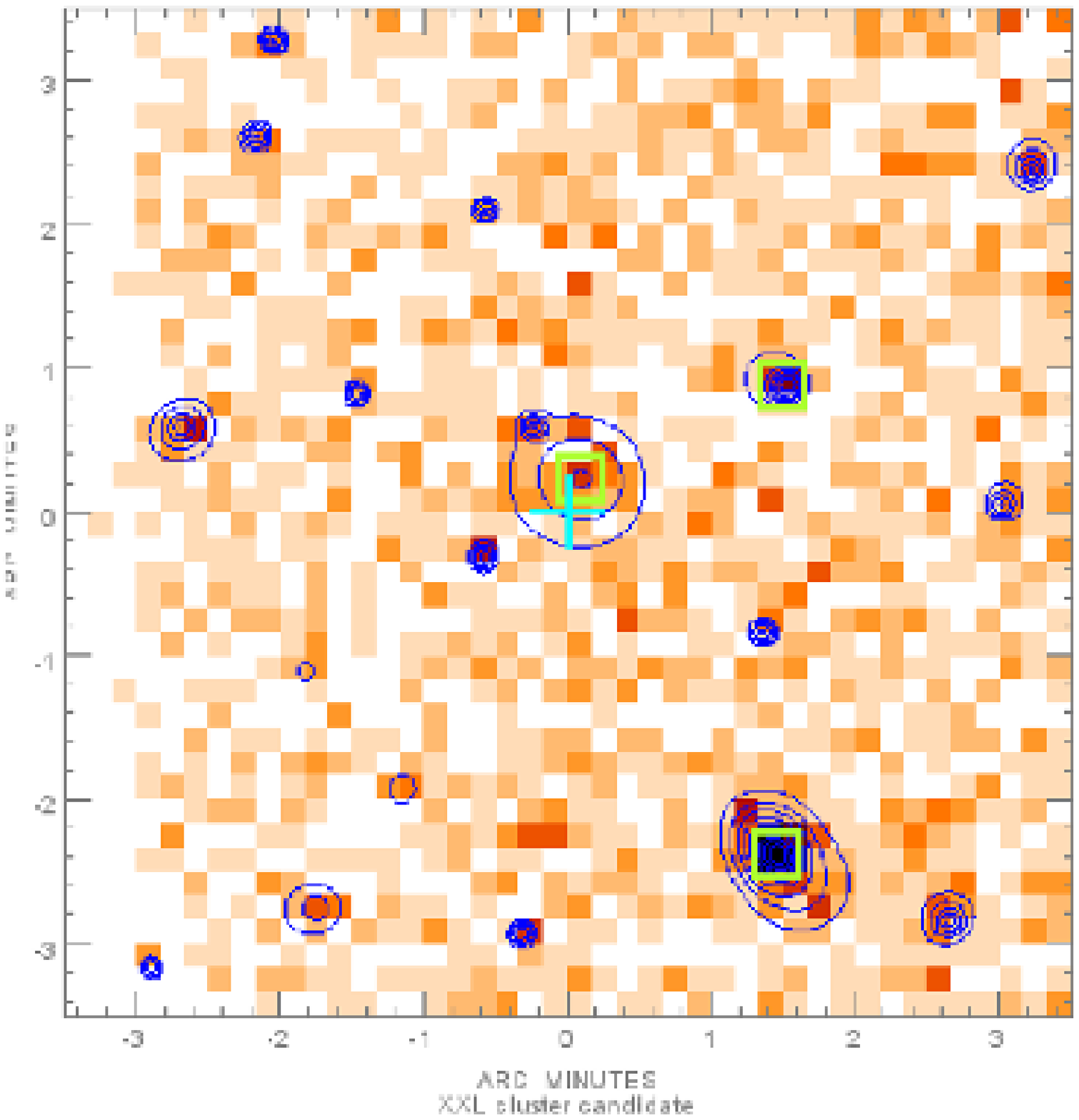}
        \end{subfigure}

        \begin{subfigure}[b]{0.33\linewidth}
                \includegraphics[width=\linewidth]{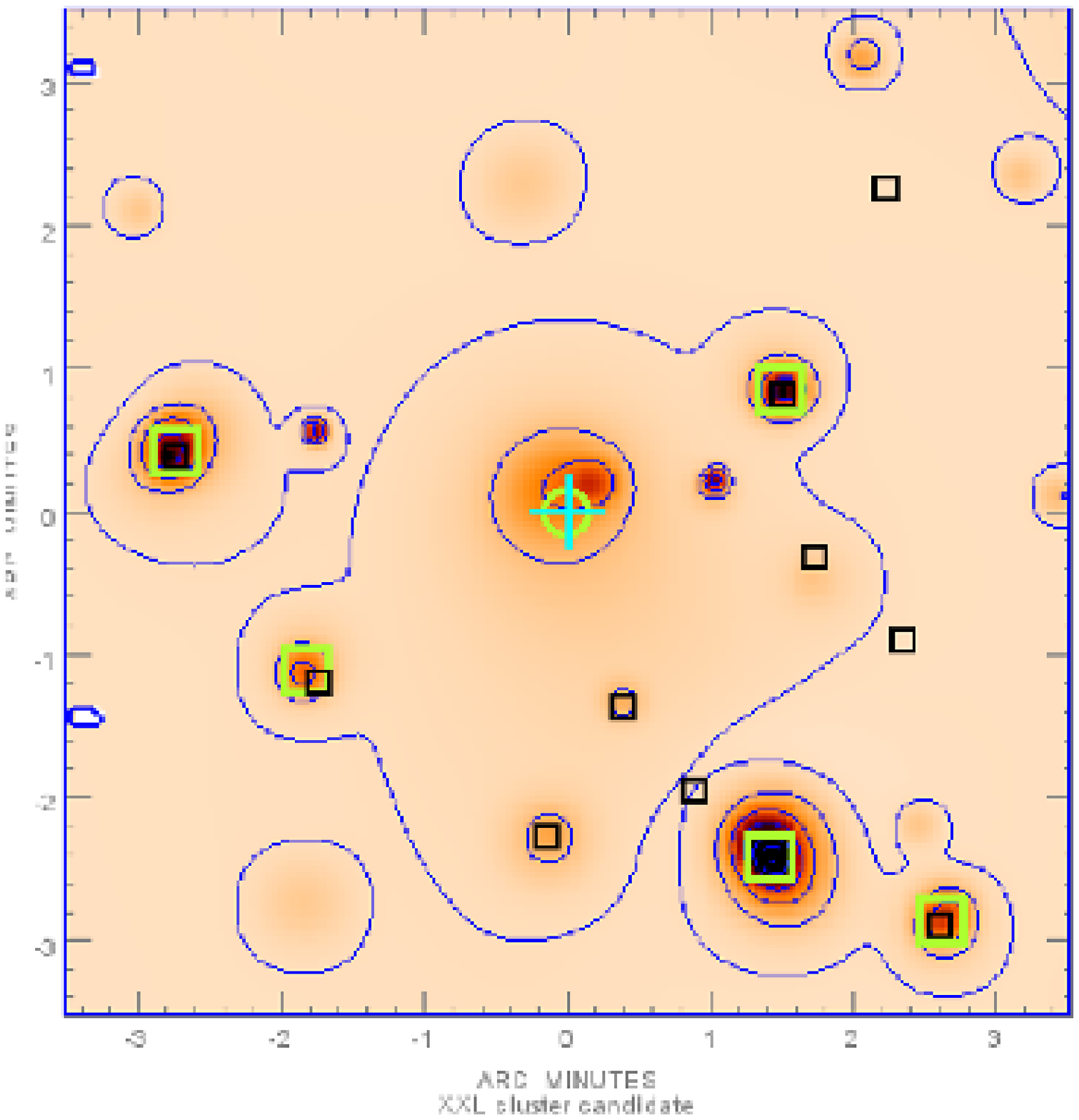}
        \end{subfigure}
        \begin{subfigure}[b]{0.33\linewidth}
                \includegraphics[width=\linewidth]{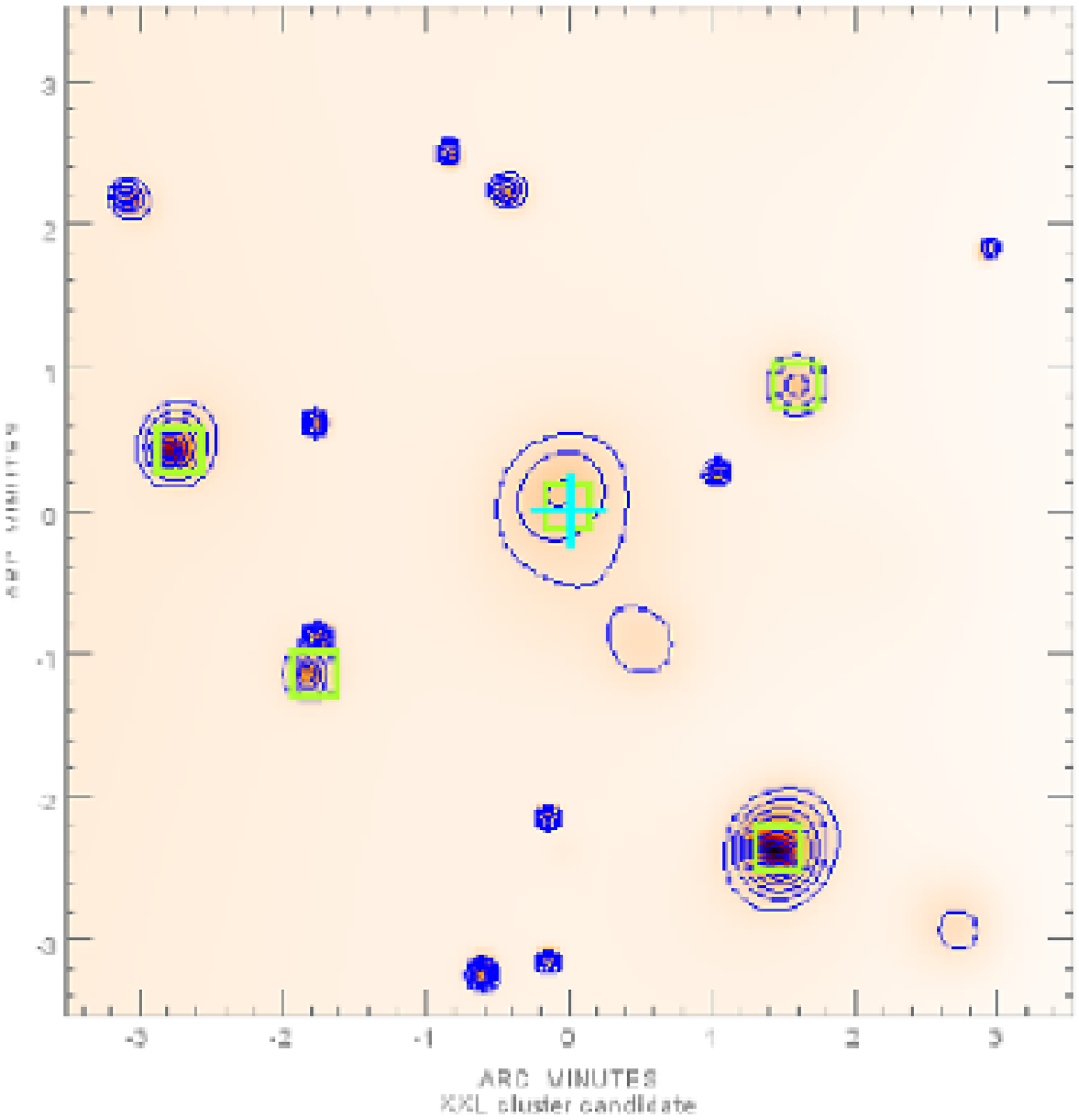}
        \end{subfigure}
        \begin{subfigure}[b]{0.33\linewidth}
                \includegraphics[width=\linewidth]{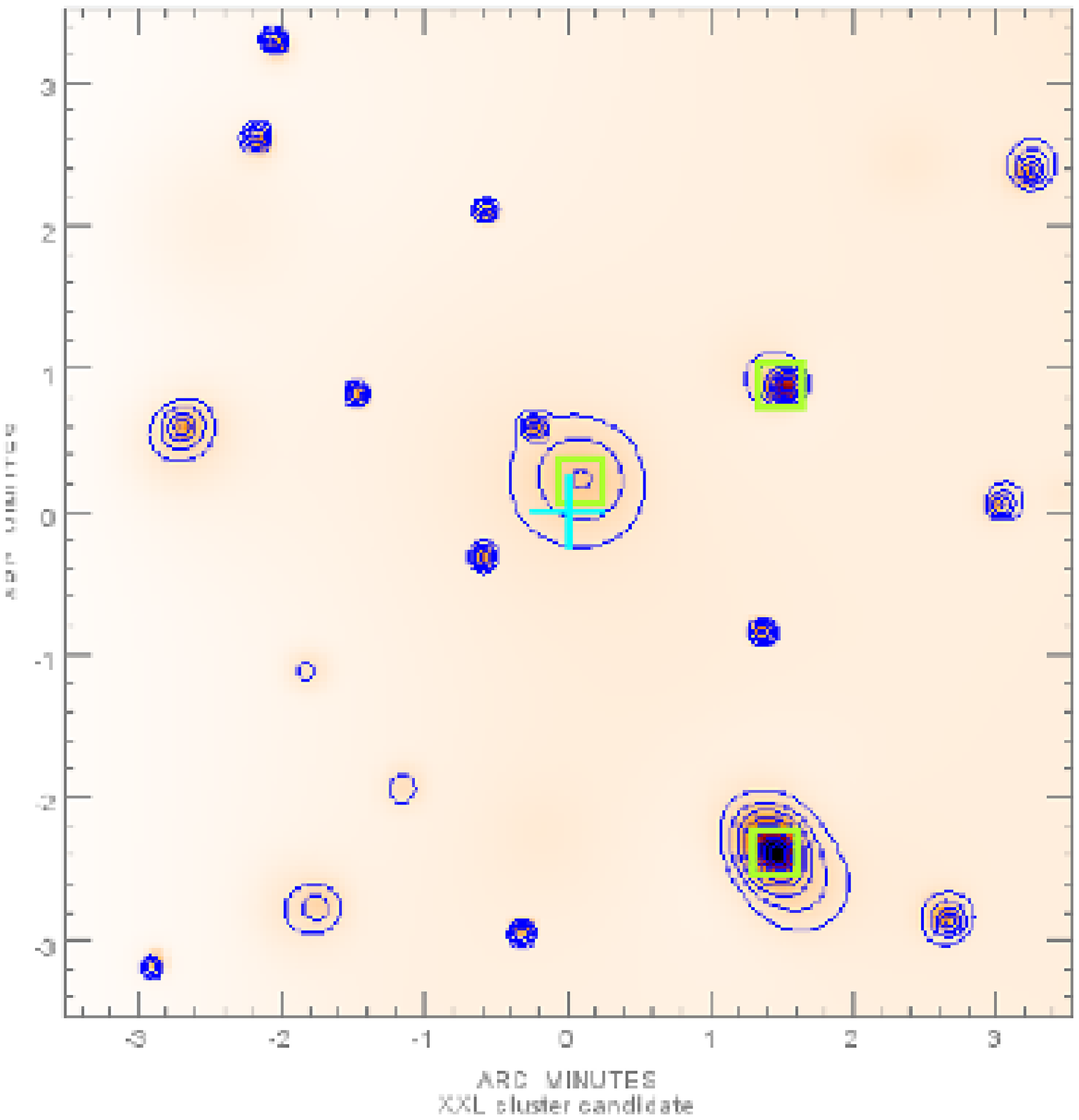}
        \end{subfigure}

        \caption{Using tiles allows us to recover a cluster that would be missed using separate pointings.
        In all images a green circle indicates an extended source (C1 or C2 detection) and a green
        square a source that is significant ($\textsc{pnt\_det\_stat}>15$) but which cannot be securely
        identified as point-like or extended; the cyan cross shows the position of the recovered
        cluster and the black squares in the left panels the position of the input AGNs; X-ray
        contours are also shown.
        Panel a (top left): combined photon image.
        Panel b (top center): photon image of one of the pointings.
        Panel c (top right): photon image of the other pointing.
        Panel d (bottom left): combined wavelet image.
        Panel e (bottom center): wavelet image of one of the pointings.
        Panel f (bottom right): wavelet image of the other pointing.
        }

        \label{fig:ov}
\end{figure*}

As an example of the power of using tiles, we show in Figure \ref{fig:ov} a
simulated weak cluster with count rate $0.01~\mathrm{count/s}$ and core radius
$10\arcsec$ observed by two pointings.
The cluster is securely recovered as a C1 using tiles, but is not recognised as such
when using single pointings separately.
In the top panel are shown, on the left, the combined photon map of the two pointings
(which is {not} used in the fit) and, at the centre and on the right, the
individual photon images of the two separate pointings.
In the bottom panel are shown the corresponding wavelet smoothed images.
As can be seen in the figure, the emission in either pointing is very weak as the
sources are at a large off-axis angle ($\approx 10\arcmin$) and are not bright enough 
to reliably identify the source as a cluster.
However, when both are used the source is securely identified as a C1 cluster
(at the position indicated by the cyan cross) with a value of
$\textsc{ext\_stat}\approx 47$, more than enough for a secure identification as a
cluster.

\subsection{Recovering clusters with central AGN contamination (AC).}
\label{sec:ac}

\begin{figure}
        \centering

        \resizebox{\hsize}{!}{\includegraphics{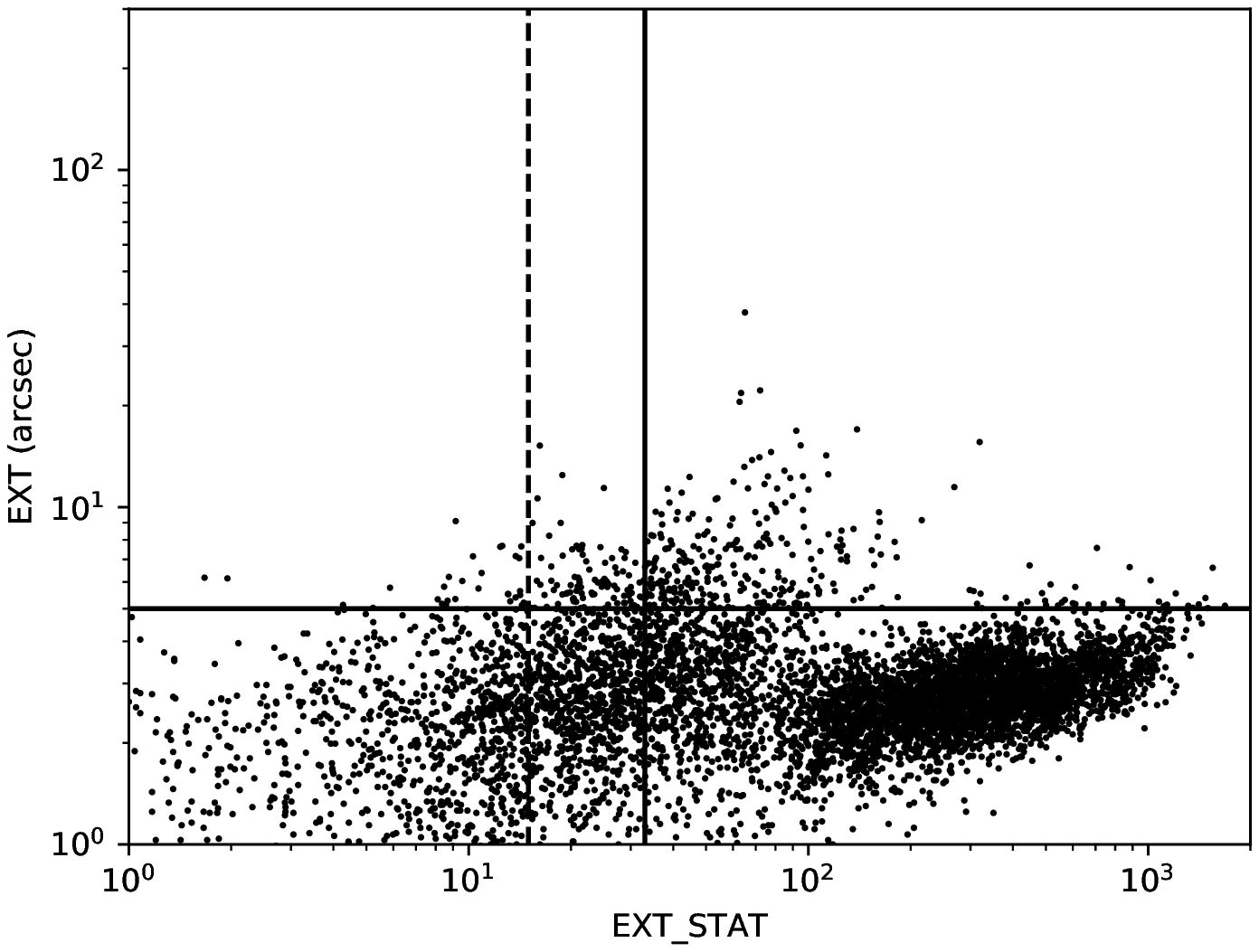}}
        \caption{\textsc{ext} vs. \textsc{ext\_stat} plane for recovered clusters contaminated
        by a central AGN (AC).
        The continuous line at $\textsc{ext\_stat}=33$ shows the cut for the C1 selection,
        and the dashed line at $\textsc{ext\_stat}=15$ shows the cut for the C2 selection.
        The figure shows that many clusters are no longer identified as C1, or even as C2,
         because of the central AGN contamination.
        }

        \label{fig:ext_vs_ext_like_ac}
\end{figure}

Figure \ref{fig:ext_vs_ext_like_ac} shows the \textsc{ext} versus \textsc{ext\_stat}
plane for recovered clusters contaminated by a central AGN where the contaminating
AGN has twice the count rate as the cluster.
The continuous line at $\textsc{ext\_stat}=33$ shows the cut for the C1 selection
and the dashed line at $\textsc{ext\_stat}=15$ shows the cut for the C2 selection.
The figure shows that many clusters are not identified as C1, or even as C2,
because of the central AGN contamination.

\begin{figure}
        \centering

        \resizebox{\hsize}{!}{\includegraphics{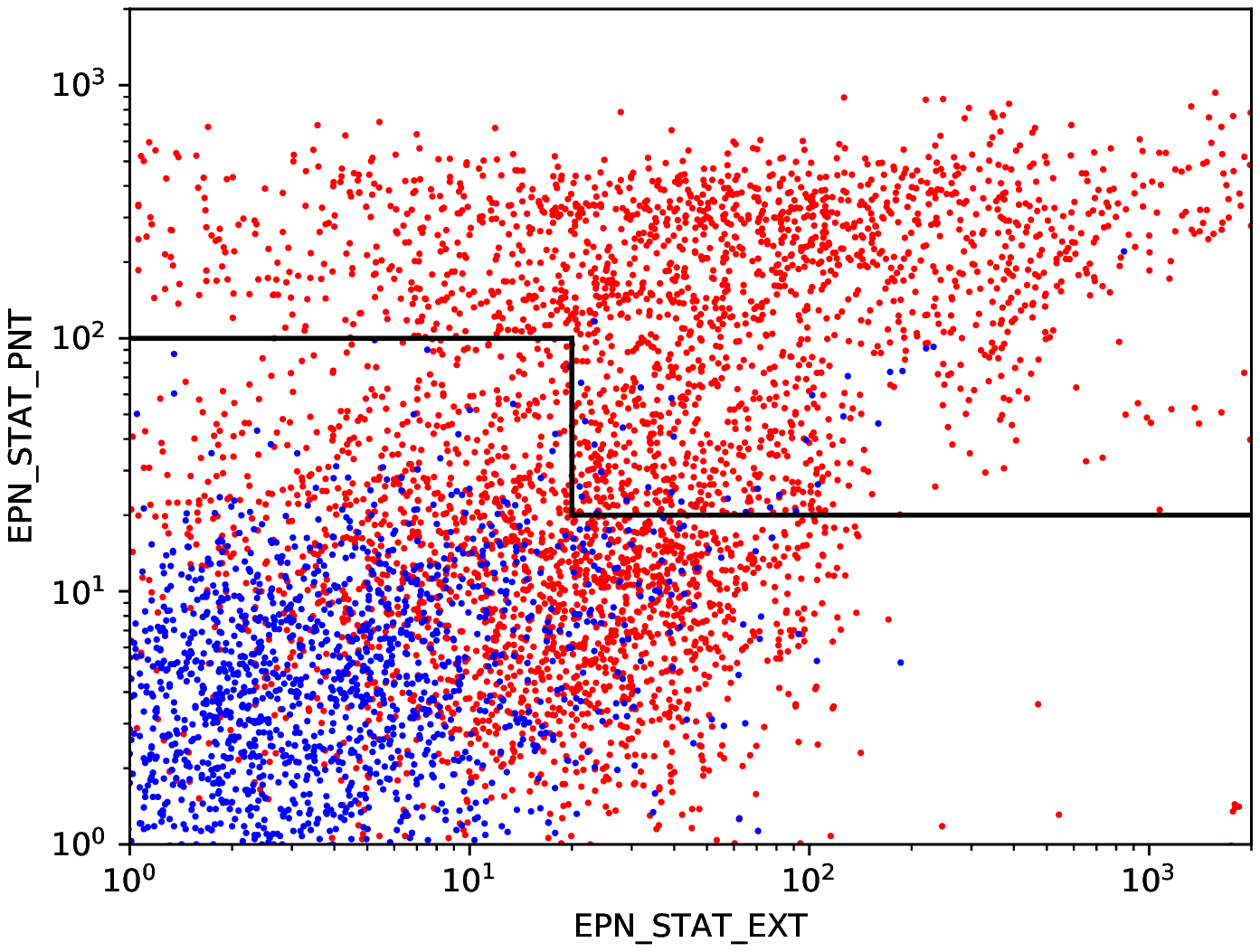}}
        \caption{
        \textsc{epn\_stat\_pnt} vs. \textsc{epn\_stat\_ext} plane for clusters
        contaminated by a central AGN (red) and AGNs (blue).
        The continuous lines  show the cuts for the AC selection.
        The figure shows that AC clusters may be recovered by applying the AC selection.
        }

        \label{fig:epn_like_pnt_vs_epn_like_ext}
\end{figure}

Figure \ref{fig:epn_like_pnt_vs_epn_like_ext} shows the \textsc{epn\_stat\_pnt} versus
\textsc{epn\_stat\_ext} plane for clusters contaminated by a central AGN (red)
and AGNs (blue).
The figure shows that the region
($\textsc{epn\_stat\_pnt}=20$ AND $\textsc{epn\_stat\_ext}=20$) OR 
($\textsc{epn\_stat\_ext}=100$) is almost empty of recovered AGNs, and can be
used to identify AC clusters.

\subsection{Distinguishing clusters from double sources.}
\label{sec:doubles}

\begin{figure}
        \centering

        \resizebox{\hsize}{!}{\includegraphics{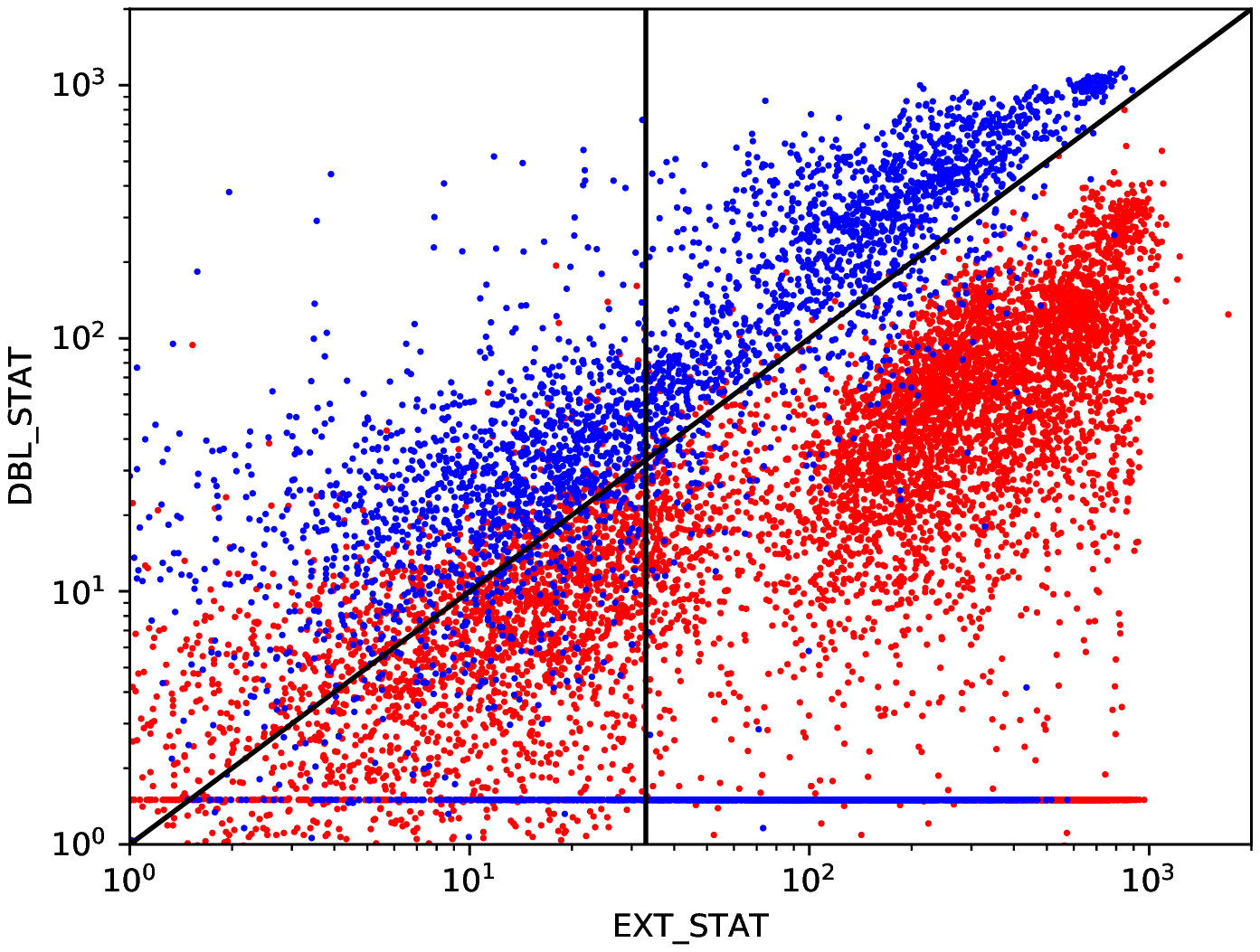}}
        \caption{\textsc{dbl\_stat} vs. \textsc{ext\_stat} plane for recovered clusters (red) and
        double sources (blue).
        The continuous line at $\textsc{ext\_stat}=33$ shows the cut for the C1 selection
        The figure shows that most double sources have $\textsc{dbl\_stat}>\textsc{ext\_stat}$.}

        \label{fig:ext_like_vs_dbl_like}
\end{figure}

We finally introduce a criterion to flag pairs of point sources appearing close in
projection on the sky (double sources or doubles); this is necessary as  point sources that are so
close (in projection) may sometimes be misidentified as an extended
source.  
These sources may be easily flagged by considering the \textsc{dbl\_stat}
statistic, defined in  Equation \ref{eq:dblstat}.
This is shown in Figure \ref{fig:ext_like_vs_dbl_like} showing the
\textsc{dbl\_stat} versus \textsc{ext\_stat} plane for recovered clusters (red) and
double sources (blue) where the continuous line at $\textsc{ext\_stat}=33$ shows
the cut for the C1 selection; the condition
$\textsc{dbl\_stat}>\textsc{ext\_stat}$
does a good job of separating these double sources from real C1s. 
The blue line at $\textsc{dbl\_stat}=1.5$ in Figure \ref{fig:ext_like_vs_dbl_like}
pertains to double sources for which $\textsc{dbl\_stat}=0$; although some of
these sources have $\textsc{ext\_stat}>33$, almost all such cases also have
\textsc{ext}$~<5\arcsec$ so there is no risk for them to be classified as C1s.

\section{ P1 selection}
\label{sec:p1}

In the same way as done for the C1/C2/AC selection, it is interesting to
investigate selection criteria for point sources that will yield samples with
high purity down to  a flux limit that is as small as possible; we refer to this new
selection as \emph{P1}.
As stated in Subsection \ref{sec:fit} a cut
$\textsc{pnt\_det\_stat}>15$ works well in distinguishing real
sources from chance background fluctuations.
As the criterion $\textsc{pnt\_det\_stat}>15$ was derived using only simulations of
point sources, there is the concern that weak, extended sources which do not pass the
C1/C2 selections but were detected with $\textsc{pnt\_det\_stat}>15$ can be
mistakenly classified as `point sources' in the sense of P06.
In the following we  refer to sources with $\textsc{pnt\_det\_stat}>15$ as
`significant detections' (SD).
We use the simulations of clusters introduced above to check the validity
of this assumption; we want to find how many input clusters in our simulations
are detected with $\textsc{pnt\_det\_stat}>15$ but do not pass the C1/C2 selection.
It might be  argued that since the number of AGNs in the sky is so much higher
than the number of clusters, possible contamination of AGN samples
from clusters flagged as SD and therefore regarded as point sources
should not be a concern.
However, this is not so as there is a correlation between clusters and AGNs:
many clusters are known to harbour a central AGN; we think it is important to have a
selection that yields a sample of point sources that is  as pure as possible, down to
 a flux limit that is as low as possible.
We have therefore introduced the class \emph{P1} for point sources with the
aim of selecting a sample of point sources, defined only in terms
of instrumental variables, that has a high degree of purity
and is complete down to a count rate that is  as low  as possible.
The P1 class is defined by the following criteria in the
$[0.5-2]~\kev$ band:

\begin{enumerate}
\item
$\textsc{pnt\_det\_stat}>30$;
\item
$\textsc{ext}<3\arcsec~\mathrm{OR}~\textsc{ext\_stat}=0$
\end{enumerate}

\begin{figure}
        \centering
        \resizebox{\hsize}{!}{\includegraphics{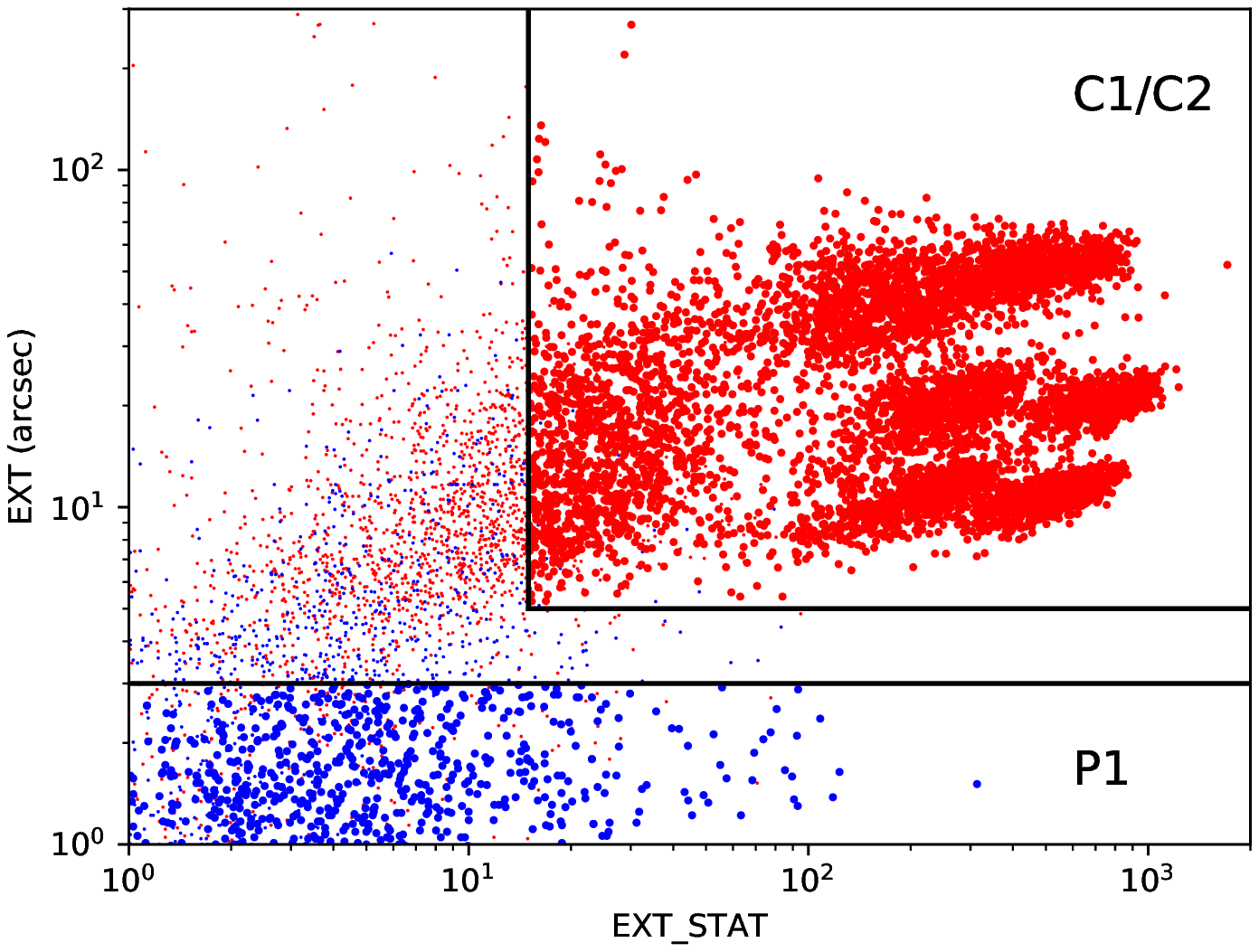}}
        \caption{$\textsc{ext}$ vs. $\textsc{ext\_stat}$ plane for recovered clusters
        (red) and AGNs (blue).
        The boxes indicate the C1/C2 and P1 selection regions, respectively.
        Sources identified as either C1/C2 or P1 are shown with larger points than
        the others.}
        \label{fig:p1_box}
\end{figure}

Figure \ref{fig:p1_box} shows the $\textsc{ext}$ versus $\textsc{ext\_stat}$ plane
recovered clusters (red) and AGNs (blue); the figure shows the C1/C2 selection box
and the P1 selection box, mutually exclusive.
Sources classified as C1/C2 or P1 are shown with larger points than the others.

\begin{table*}
\caption{Simulated clusters recovered as P1. Total input clusters: 784.}
\label{tab:p1}
\centering
\begin{tabular}{lcrl}

\hline
Input count rate    & Input core radius & P1 Detections & P1 Fraction \\
$\mathrm{count/s}$  & $\arcsec$         &               &             \\
\hline
0.005            & 10 &  38 &  0.048 \\
0.005            & 20 &  36 &  0.046 \\

0.01             & 10 &  57 &  0.072 \\
0.01             & 20 &  20 &  0.025 \\

0.05             & 10 &  7 &  0.009 \\
0.05             & 20 &  0 &  0.0   \\
0.05             & 40 &  2 &  0.025 \\

0.1              & 10 &  7 &  0.009 \\
0.1              & 20 &  1 &  0.001 \\
0.1              & 50 &  0 &  0.0   \\

\hline
\end{tabular}
\end{table*}

\begin{figure}
\resizebox{\hsize}{!}{\includegraphics{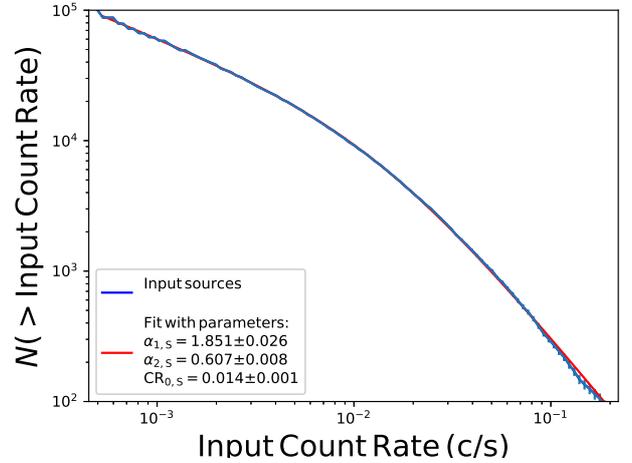}}
\caption{$\log(N(>$$CR))-\log(CR)$ of input simulated point sources.}
\label{fig:input_pointsources}
\end{figure}

Table \ref{tab:p1} shows that for the P1 selection the fraction of input
clusters that are classified as P1 sources is quite low ($\approx 3\%$ or less)
for almost all combinations of count rates and core radii;
therefore, the P1 selection
we have introduced is expected to give virtually pure samples of point sources,   first because
AGNs greatly outnumber clusters, and second because a small fraction
of clusters is mistakenly classified as P1 for any realistic combination of count rates
and core radii.
To test the validity of the P1 selection criteria we have simulated four patches of sky of
$5\deg\times 5\deg$ each containing only randomly distributed AGNs; each $5\deg\times 5\deg$
patch of sky contains $\approx 80000$ AGNs with flux distribution from \citet{moretti03} and
down to a flux limit $10^{-16}~\funits$, exactly as done for the other simulations.
Figure \ref{fig:input_pointsources} shows the cumulative distribution of the
input AGNs, ($\log(N($$>CR))-\log(CR)$, with

\begin{equation}
\label{eq:moretti}
N(>CR)\propto\frac{1}{CR^{\alpha_{1,S} }+CR_{0,S}{}^{\alpha_{1,S}-\alpha_{2,S}}\times CR^{\alpha_{2,S}}},
\end{equation}
with $\alpha_{1,CR}=1.851$, $\alpha_{2,CR}=0.607$, and $CR_{0,CR}=0.014~\mathrm{count/s}$,
appropriate for the $[0.5-2]~\kev$ band, from
\citet{moretti03}.\footnote{\citet{moretti03} work with physical fluxes; we work with
instrumental count rates and adopt a conversion factor
$1.086\times 10^{-12}\funits / (\mathrm{count/s})$, computed from
WebSpec and appropriate for a power law spectral model with index $1.7$ and THIN XMM filters.}

\begin{figure*}
        \centering

        \begin{subfigure}[b]{0.30\linewidth}
                \includegraphics[width=\linewidth]{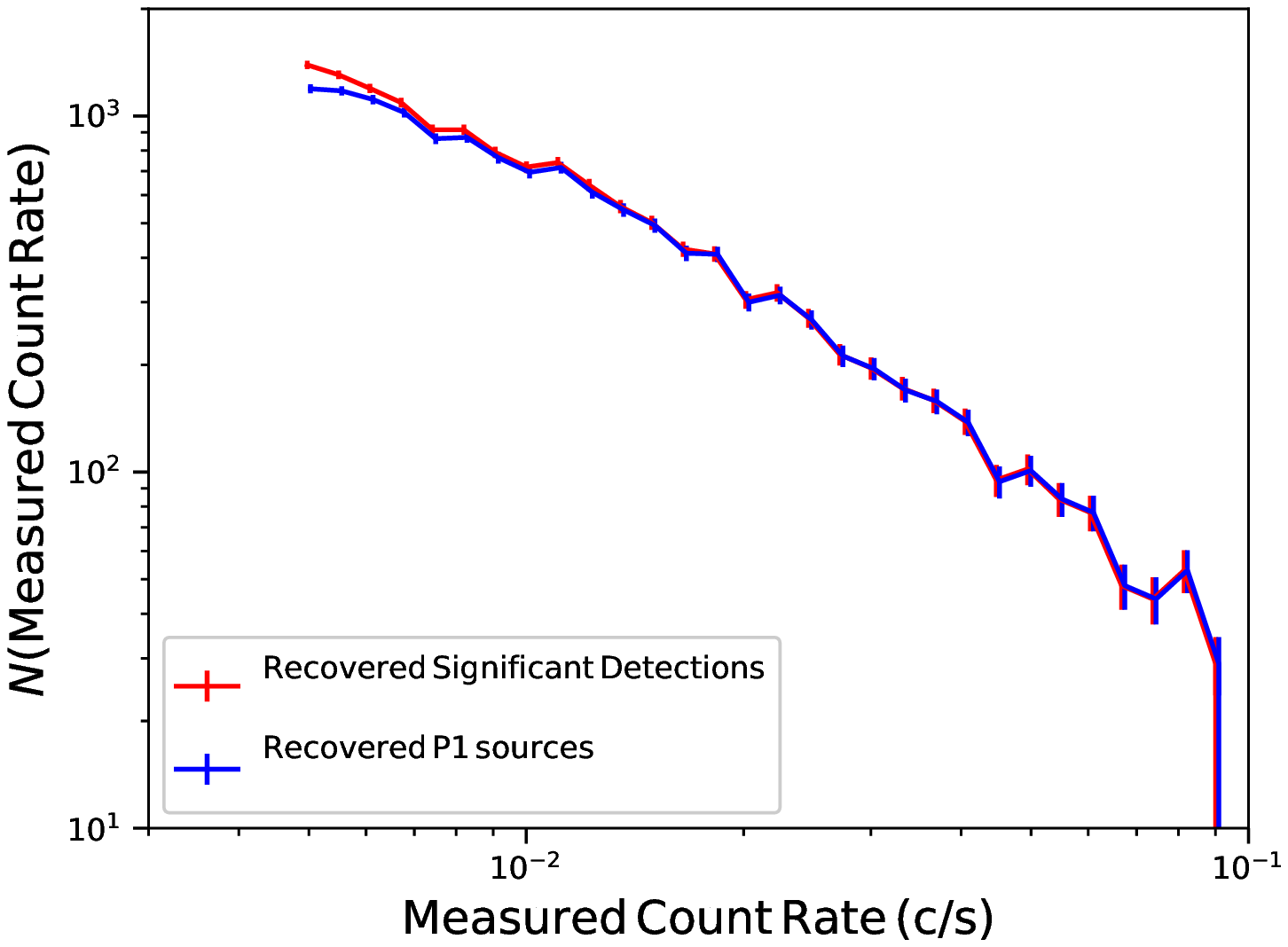}
        \end{subfigure}
        \begin{subfigure}[b]{0.30\linewidth}
                \includegraphics[width=\linewidth]{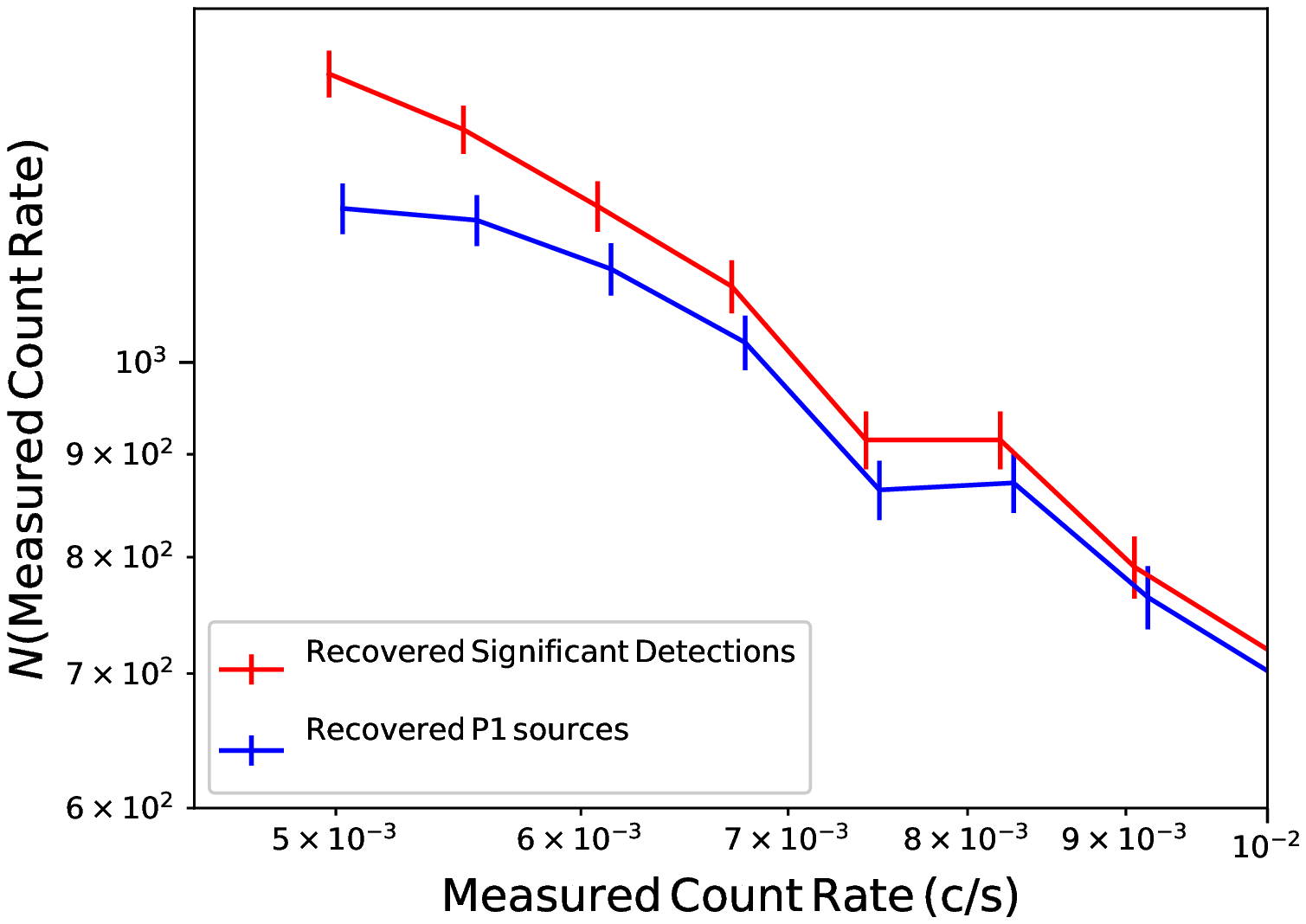}
        \end{subfigure}

        \caption{Differential $\log(N(CR))-\log(CR)$ for recovered sources.
        Blue line: all significant detections;
        Red line: P1 detections.
        The figure shows that the P1 selection is complete down to $CR \approx 0.006~\mathrm{count/s}$.
        Panel a (left): overall view.
        Panel b (right): blow up of the regime where P1 selection becomes incomplete.
        }

        \label{fig:recovered_pointsources}
\end{figure*}

Figure \ref{fig:recovered_pointsources} shows the differential distribution (non-cumulative)
$\log(N(CR)-\log(CR)$) both for significant detections (\textsc{pnt\_det\_stat}$>15$) and for
P1 sources.
The left panel shows that the P1 selection is virtually complete down to 
to $CR \approx 0.006~\mathrm{count/s}$; the right panel shows a blow up of the
plot at $CR \approx 0.006~\mathrm{count/s}$ making it clearer.
Table \ref{tab:sources} lists the selection criteria we have introduced.

\begin{table*}
\caption{Summary of the various types of source and their selection criteria.
If more than one condition is specified, all
conditions must be used unless explicitly stated otherwise.
}
\label{tab:sources}
\centering
\begin{tabular}{clll}

\hline
Classification  & Source type   & Selection criteria                                        & Note \\
\hline
C1              & Extended      & $\textsc{ext}>5\arcsec$                                   & Negligible contamination. \\
                &               & $\textsc{ext\_stat}>33$                                   & \\
                &               & $\textsc{ext\_det\_stat}>32$                              & \\
\hline
C2              & Extended      & $\textsc{ext}>5\arcsec$                                   & \\
                &               & $\textsc{ext\_stat}>15$                                   & \\
\hline
AC              & Extended$+$   & $\textsc{epn\_ext}>5\arcsec$                              & \\
                & central point & ($\textsc{ext\_stat\_ext}>20~\mathrm{AND}$                & \\
                &               & $\textsc{ext\_stat\_pnt}>20$)                             & \\
                &               & $\mathrm{OR}~(\textsc{ext\_stat\_pnt}>100)$               & \\

\hline
P1              & Point         & $\textsc{pnt\_det\_stat}>30$                              & Negligible contamination from clusters.\\
                &               & $\textsc{ext}<3\arcsec~\mathrm{OR}~\textsc{ext\_stat}=0$  & Complete down to $\approx 0.006~\mathrm{count/s}$. \\
\hline

                & Double        & $\textsc{dbl\_stat}>\textsc{ext\_stat}$                   & Auxiliary cut to decide dubious \\
                &               &                                                           & cases during human screening. \\
\hline
\end{tabular}
\end{table*}

\section{Conclusions}
\label{sec:conclusion}

We presented \xaminF, the newest and final pipeline for the XXL
survey, we detailed the main improvements with respect to the older pipeline \xaminP,
and we validated it through an extensive set of simulations.
The need for an updated version, even though \xaminP~already
works very well, stems mostly from the need to make optimal use of the
available X-ray information, using multiple observations of the same
source when available.

\par
Using \atile~tiles for source detection and all available observations
when fitting a candidate source we showed
that we can significantly increase the number of detected clusters;
clusters that are too faint in either observations and would be missed
(or found as C2 at best) can be reliably found. This is particularly
important for the low-mass, low-luminosity systems which are the bulk
of the XXL population.

\par
Another major improvement we introduced is the capability to detect clusters
contaminated by a central AGN by simultaneously fitting for an extended
and a point source.
The presence of an AGN at the centre of a cluster
severely degrades the performance of the simple extended source fit, causing
the cluster to be missed in many cases, especially where the AGN count rate
is close to or higher than the cluster rate.
We have shown that the introduction of the \textsc{epn} fit allows us to recover the
contaminated cluster in many cases.

\par
We also introduced the capability of fitting two point
sources, useful for flagging pairs of AGNs seen close in projection that
may otherwise be incorrectly identified as a cluster; we were able to
define, through simulations, a criterion for flagging double sources
based again on purely instrumental variables.

\par
Although we have shown that \xaminF~is an excellent pipeline for the XXL survey, and
in principle could be adapted to other X-ray surveys, more work remains to be done.
In an upcoming paper we will test the \xaminF~performance using more
realistic hydrodynamic simulations where the shape of clusters is not circular and using
a more realistic population of AGNs; we will also consider different exposure times
and the consequences of adopting different values of $\beta$.
All this will be necessary to allow the XXL survey to achieve its ultimate goal
of precision cosmology with X-ray selected clusters of galaxies.

\begin{appendix}
\section{Summary of simulations of different types of sources}
\label{sec:app}

A summary of all the simulations of clusters used is given in Table \ref{tab:simc}.
A summary of all the simulations of clusters with a central AGN used is given in Table \ref{tab:simcc}.
A summary of all the simulations of double sources used is given in Table \ref{tab:simd}.

\begin{table*}
\caption{Summary of simulated clusters.}
\label{tab:simc}
\centering
\begin{tabular}{lcccc}

\hline
\multicolumn{5}{c}{$\beta=2/3$, exposure time $10~\mathrm{ks}$.} \\
\hline
Count rate ($\mathrm{count/s}$) & Core radius ($\arcsec$) & Number of clusters & Number of clusters         & Number of clusters \\
           &                       &                      & far from image borders                          & at $>10\arcmin$ from \\
           &                       &                      & and used in computing                           & all pointings \\
           &                       &                      & detection probabilities                         & \\
\hline

0.005            & 10 & 841 & 784 & 548 \\
0.005            & 20 & 841 & 784 & 548 \\
0.005            & 50 & 841 & 784 & 548 \\

0.01             & 10 & 841 & 784 & 548 \\
0.01             & 20 & 841 & 784 & 548 \\
0.01             & 50 & 841 & 784 & 548 \\

0.05             & 10 & 841 & 784 & 548 \\
0.05             & 20 & 841 & 784 & 548 \\
0.05             & 40 & 841 & 784 & 548 \\
0.05             & 50 & 841 & 784 & 548 \\

0.1              & 10 & 841 & 784 & 548 \\
0.1              & 20 & 841 & 784 & 548 \\
0.1              & 50 & 841 & 784 & 548 \\

\hline
\end{tabular}
\end{table*}

\begin{table*}
\caption{Summary of simulated clusters with central AGN.}
\label{tab:simcc}
\centering
\begin{tabular}{lclc}

\hline
\multicolumn{4}{c}{$\beta=2/3$, exposure time 10ks.} \\
\hline
Count rate ($\mathrm{count/s}$) & Core radius ($\arcsec$) & AGN count rate ($\mathrm{count/s}$) & Number of clusters \\
\hline

0.005              & 10 & 0.01 & 841 \\
0.005              & 20 & 0.01 & 841 \\
0.005              & 50 & 0.01 & 841 \\

0.01               & 10 & 0.02 & 841 \\
0.01               & 20 & 0.02 & 841 \\
0.01               & 50 & 0.02 & 841 \\

0.05               & 10 & 0.1 & 841 \\
0.05               & 20 & 0.1 & 841 \\
0.05               & 50 & 0.1 & 841 \\

0.1                & 10 & 0.2 & 841 \\
0.1                & 20 & 0.2 & 841 \\
0.1                & 50 & 0.2 & 841 \\

\hline
\end{tabular}
\end{table*}

\begin{table*}
\caption{Summary of simulated double sources.}
\label{tab:simd}
\centering
\begin{tabular}{lccc}

\hline
\multicolumn{4}{c}{Exposure time 10ks.} \\
\hline
Count rate source 1 ($\mathrm{count/s}$) & separation($\arcsec$) & Count rate source 2  & Number of double sources \\
                                         &                       & relative to source 1 & \\
\hline

0.005              &  6 & 1 & 841 \\
0.005              & 12 & 1 & 841 \\

0.01               &  6 & 1 & 841 \\
0.01               & 12 & 1 & 841 \\

0.05               &  6 & 1 & 841 \\
0.05               & 12 & 1 & 841 \\

0.1                &  6 & 1 & 841 \\
0.1                & 12 & 1 & 841 \\

\hline
\end{tabular}
\end{table*}

\end{appendix}

\begin{acknowledgements}

XXL is an international project based around an XMM Very Large Programme surveying two  $25\desqg$
extra-galactic fields at a depth of $\approx 5\times10^{-15}\funits$ in the $[0.5-2]~\kev$ band for point-like sources.
The XXL website is http://irfu.cea.fr/xxl.
Multi-band information and spectroscopic follow-up of the X-ray sources are obtained through a number of survey programmes,
summarised at http://xxlmultiwave.pbworks.com/.

We thank Ian McCarthy for making available the X-ray fluxes of the input halos in the cosmoOWLS
simulation.  

The Saclay group acknowledges long-term support from the Centre National d'Etudes Spatiales (CNES).

LF thanks CEA for a long-term research contract.

EK thanks CNES and CNRS for support during his post-doctoral research.

We thank Jean-Luc Starck for supplying us with an improved version of \texttt{MR/1}
from which our IDL implementation was derived.

\end{acknowledgements}

\bibliographystyle{aa}
\bibliography{32931_corr}{}

\end{document}